\documentclass[11pt]{article}
	\pdfoutput=1
\usepackage[T1]{fontenc}

\usepackage{listings}
\lstnewenvironment{arkady}  
{\lstset{language=C,frame=trbl,basicstyle = \footnotesize \ttfamily , breaklines = true,showstringspaces=false}}{}

\usepackage{datetime}
\usepackage{comment}					

\usepackage[usenames,dvipsnames]{xcolor}

\usepackage[Euler]{upgreek}

\usepackage[parsep]{collref}	

\usepackage{amsmath, amssymb,amsthm}
\usepackage{stackrel}
\numberwithin{equation}{section}
\usepackage{bm,environ,mathrsfs,array,arydshln}
\usepackage{booktabs,float,slashed}
\usepackage{appendix}
\usepackage[mathcal]{euscript}
\usepackage{tensor} 						
\usepackage{mathabx}
\usepackage[vcentermath]{youngtab}
\usepackage{simpler-wick}

\usepackage{graphicx,epsfig,epic}
\usepackage[setpagesize=false,pagebackref=false, 
linktocpage, bookmarksopen=true, colorlinks=true, 
linkcolor=Maroon,citecolor=Maroon,urlcolor=Maroon]{hyperref}

\allowdisplaybreaks

\usepackage{framed}						
\definecolor{shadecolor}{rgb}{0.9996078, 0.984314, 0.960784}
\definecolor{framecolor}{rgb}{0,0,0}
\definecolor{TFTitleColor}{RGB}{1,1,1}

\newenvironment{frshaded}{%
    \MakeFramed {\FrameRestore}}%
    {\endMakeFramed}

\definecolor{myred}{RGB}{233, 33, 45}

\newcommand{\bs}{\begin{frshaded}}			
\newcommand{\es}{\end{frshaded}\noindent}

\def\ba#1\ea{\begin{align}#1\end{align}}		        
\newcommand{\be}{\begin{equation}}
\newcommand{\ee}{\end{equation}}
\newcommand{\bea}{\begin{equation} \begin{aligned}} 
\newcommand{\eea}{\end{aligned} \end{equation}}
\newcommand{\mc}{\mathcal }

\newcommand{\mk}{\mathfrak}
\newcommand{\la}{\label}
\newcommand{\eps}{\varepsilon}

\newcommand{\lp}{\notag \\ & }

\DeclareMathOperator{\Tr}{\text{Tr}}
\DeclareMathOperator{\tr}{\text{tr}}
\DeclareMathOperator{\vol}{vol}

\def \AdS  {{\rm AdS}}
\def \adst {AdS$_3$}  \def \ZZ  {{\mathbb Z}}

\newcommand{\sql}{\sqrt\l}
\renewcommand{\l}{\lambda}

\newcommand{\adstwo}{AdS$_{2}$ }

\newcommand{\gs}{{\rm G}}
\newcommand{\G}{\Gamma}
\newcommand{\w}{{\rm w}}

\def \np {\newpage}
\def \ed {\np \small
\baselineskip 11pt
\bibliography{m2s}
\small
\bibliographystyle{JHEP-v2.9}
\end{document}
}
\def \iffa  {\iffalse}
\def \te {\textstyle}
\newcommand{\rf}[1]{(\ref{#1})}
\def\ov{\over}
\def \ci {\cite}
\def \foot {\footnote}

\def\la{\label}\def \a {\alpha}
\def\foot{\footnote}
\def \adss {AdS$_5 \times S^5$\ }

\def \tb {$T\bar T$ }
\def \s {\sigma} \def \del {\partial} 
\def \ha {\tfrac{1}{2}}
\def \DD {{\rm D}}

\def \OO {{\cal O}}
\def \cc {{\rm c}}
\def \no {\nonumber}
 \def \g {\gamma}
 
\newcommand{\ad}{\text{ad}}

\newcommand{\hyp}{\mathrm{H}}

\newcommand{\dd}{\dot{d}}

\newcommand{\CP}{\mathbb{CP}}
\newcommand{\ii}{\mathrm i}
\newcommand{\cL}{\mathcal L}
\newcommand{\cM}{\mathcal M}
\newcommand{\cO}{\mathcal O}
\newcommand{\cX}{\mathscr X}

\newcommand{\al}[1]{\begin{align}{#1}\end{align}}

\begin{document}
\begin{titlepage}
\begin{tabbing}
\hspace*{10.5cm} \=  \kill 
\>  Imperial--TP--2026--SK--02 \\
\> 
\end{tabbing}



\vspace*{05 mm}
\begin{center}
{\Large\sc  \bf     
M2 brane in $\AdS_4\times S^7/\ZZ_k$:  2-loop correction to   \\[2mm]
 $\frac12$-BPS Wilson loop vs  localization in ABJM theory }

\vspace*{07mm}

M. Beccaria$^{a}$, \ \ S.A. Kurlyand$^{b}$,\ \ \ A.A. Tseytlin$^{b,}$\footnote{Also at  ITMP of MSU  and Lebedev Inst.},
\ \ J. van Muiden$^{b,}$\footnote{\texttt{matteo.beccaria@le.infn.it,\ s.kurlyand23@ic.ac.uk, \ tseytlin@ic.ac.uk, \ jvanmuid@ic.ac.uk}}

\vspace*{4mm}
{\small
	
${}^a$ Universit\`a del Salento, Dipartimento di Matematica e Fisica \textit{Ennio De Giorgi},\\ 
		and INFN - sezione di Lecce, Via Arnesano, I-73100 Lecce, Italy
			\vskip 0.3cm
${}^b$ Abdus Salam Centre for Theoretical Physics,\\ Imperial College London,  SW7 2AZ, U.K.
			\vskip 0.01cm
}
\vspace*{1.2cm}

\end{center}

\begin{abstract} 
We study the quantization of an M2 brane wrapping
$\mathrm{AdS}_2\times S^1$ inside
$\mathrm{AdS}_4\times S^7/\mathbb{Z}_k$, dual to the
$\frac{1}{2}$-BPS circular Wilson loop in ABJM theory.
The classical action and the one-loop contribution were previously shown
to reproduce the gauge-theory localization result.
Here we extend the analysis to the  two-loop order.
In an adapted static and $\kappa$-symmetry gauge the world-volume action
contains no cubic interaction vertices, so that the two-loop correction is
determined entirely by quartic interactions and is manifestly free of
logarithmic UV divergences.
We find that the bosonic and fermionic contributions combine into a
perfect square $ {\mathscr X}^2$   where ${\mathscr X}$ is a linear combination of the bosonic and fermionic 
3d Green's functions and their derivatives. 
The evaluation of $\mathscr X$   requires fixing the $\mathrm{AdS}_2$ boundary quantization
of the massless fermionic modes.
With the boundary conditions selected by the preserved supersymmetry,
$\mathscr X$  vanishes and hence the complete two-loop correction is found to  be  zero.
This result would not  be consistent with  identifying the M2-brane loop expansion
directly with the large-$N$ expansion of the canonical, fixed-$(N,k)$
localization prediction.
Instead, it agrees precisely with the recent proposal that the
semiclassical M2-brane partition function should be matched to the
gauge-theory grand-canonical ensemble, in which the Wilson loop expectation
value is perturbatively one-loop exact.
The vanishing of the two-loop correction thus provides a non-trivial test of
this grand-canonical identification.
\end{abstract}

\vskip 3.5cm
\end{titlepage}

\def \adst {AdS$_2$ }
\def \ept {\eps'}
\def \adstwo {\adst}
  \def \rmb {{_{\rm b}}}   \def \rmf {{_{\rm f}}}  \def \rmbf {{_{\rm bf}}}
 \def \DD {{\mk D}}  \def \b {\beta } 
\def \m {\mu} \def \n {\nu} \def \L {\mc L}
 \def \ep {\eps}
  \def \ads {{\rm AdS}} \def \four {\tfrac{1}{ 4}}
\def \fo {\tfrac{1}{4}} \def \beg {\be}
\def \F  {F} \def \RR {{\Lambda_{\rm IR}}} \def \TT {{\rm T}}
\def \DD {{\rm D}}
\def \TT  {{\rm T} } 
\def \ep {\epsilon}
\def \vd {\dot v}   \def \ad {\dot a} \def \zb {\bar z} 
\def \RR {{\bar \Lambda }} 

\def \iffa {\iffalse} \def \aa {{\rm a }}  
\def \Tr {{\rm Tr}}
\def \tr {{\rm tr}}
\def \vol   {{\rm vol}}
\def \rr {{\rm r}}
\def \OO {{\cal O}}
\def \cc {{\rm c}}
\def \kk {\kappa}\def \ka {\kappa} \def \L  {\Lambda}
\def \rT {{\rm T}}
\def \tt {{\rm t}}
\def \half {\tfrac{1}{2}}
\def  \adsss  {AdS$_7\times S^4\ $}
\def  \adssp  {AdS$_5\times S^5\ $}
\def \adss {AdS$_7\ $}
\def \adst   {AdS$_3\ $}
\def \N   {{\cal N}}
\def \half {\tfrac{1}{2}}
\def \xx {{ x}}
\def \vx {\vec{x}}
\def \fo {{1\ov 4}}\def \n {\nu}
\def \la {\label} \def \fo {{1\ov 4}}
\def \beg  {\begin{equation}}
\def \eeg {\end{equation}}
\def \tet {\textstyle}
\def \lag {\langle} \def \rag {\rangle}
\def \xxx {z}
\def \xxxx {{\rm z}}
\def \ze {\chi}
\def \zz {{v}}
\def \adsp {AdS$_5 \times S^5$}
\def \rh   {\hat {\rm G}}
\def \rG {{\rm G}}
\def \rT {{\rm T}} 
\def \ssss   { {\hat {\rm G}}_\theta}
\def \z  {\zeta} 
\def \rhg {\tilde {\rm G}}
\def \w  {\s}
\def \halpha {{\hat \alpha}}  \def \hbeta {{\hat \beta}} \def \hgamma {{\hat \gamma}}
\def \K {{\rm K}}
\def \gs  {\rG}   \def \rh {\hat \rG}
\def \tgt {\tilde {\rm G}_\theta}
\def \bgt {\bar {\rm G}_\theta} 
\def \La  {\Lambda}  \def \tG {\G_\theta}  
\def \hyp {{\rm H}} \def \ad {{\rm ad}\, }
\def \onevec  {{\bf 1}}
\newcommand{\Gb}{\bar\Gamma}
\def \vthree {}  
\def \vfour  {}  
\def \vfive  {}  

\newcommand{\Gk}{\Gamma_\kappa}
\def \sgn {{\rm sgn}}
\newcommand{\Dslash}{\slashed\nabla}

\newcommand{\rmi}{i}
\newcommand{\rme}{\mathrm{e}}
\newcommand{\rmd}{\mathrm{d}}

{\footnotesize
\makeatletter
\newcommand*{\toccontents}{\@starttoc{toc}}
\makeatother
\toccontents
}
\def \bT {\bar{\rm T}}
\def \TT  {{\bT}_2}
\def \lp {\ell_{\rm p}} 
\def \NN {{\rm N}}
\def \dd {d}
 \def \rT {{\rm T}} 
  \def \adss {$\AdS_4 \times S^7/\ZZ_k\ $}
\def \us {\upsigma}
\def \RR {{\rm R}}  \def \ls {\ell_{s}}
\def \L {{\cal L}}\def \cL {{\cal L}}   \def \G {{\rm G}}
\def \X {{\rm X}} 
\def \vT {{\rm t}}
\def \mC {\cX_b}
\def \tb {\bar \theta}
\def \mR {\cX_f}
\def \MM {{\rm M}}
\def \T {{\rm t}}
\def \KK {{\cal K}}
\def \ss {{\uptau}}
\def \PPi {{\mathcal P}}
\def \GG  {\hat \G}
\def \uN  {\upgamma} \def \Gba  {\bar \gamma}
\def \HH  {{\rm H}}
\def \SS  {{\rm S}}
\def \rZ  {{\rm Z}}

\newcommand{\jesse}[1]{\textbf{\textcolor{blue}{[JvM:~#1]}}}

\setcounter{footnote}{0}

\section{Introduction and summary}\label{s1}

Quantum M2  branes in 11d  space  provide a tool   to study  AdS/CFT   beyond  the planar limit. The prime  example  is M-theory in   $\AdS_4\times S^7/\ZZ_k$ dual to ABJM theory \ci{Aharony:2008ug}. 

In particular, localization \ci{Kapustin:2009kz,Drukker:2010nc}  provides a prediction for the part  perturbative in $1/N$ of the  expectation value of  the  $1\ov 2$-BPS  circular Wilson loop  (WL) \cite{Klemm:2012ii}.
At fixed $N$ and $k>2$  it is given by  (modulo  corrections non-perturbative at large $N$)
\begin{align}
 \langle W(N,k)\rangle
 =\frac{1}{2\sin{2\pi\ov k}}
 \frac{\rZ(N-{2\ov k},k)}{\rZ(N,k)}, \la{1}
 \qquad \qquad  
 \rZ(N,k)
 =C^{-1/3} \rme^{A} \operatorname{Ai}\!\big[C^{-1/3}(N-B)\big],
\end{align}
where $C=\frac{2}{\pi^2k},
 \ 
 B=\frac{k}{24}+\frac{1}{3k}
 $  
 and $A(k)$ is a ``constant-map''  contribution  that cancels out in the ratio.    
  The above expression  is valid for $k>2$, the case that we will be considering below. 
Expanded in  large $N$  for fixed level $k$ this gives 
\begin{equation}
 \langle W(N,k)\rangle
 =\frac{1}{2\sin{ 2\pi \ov k}} \, e^{\pi\sqrt{2N/k}} 
 \Big[1-\frac{\pi (k^2 + 32) }{24 \sqrt 2  k^{3/2} } {1 \ov \sqrt{N}}
 +\OO(N^{-1})\Big]\ .
 \label{3}
\end{equation}
 $ \langle W\rangle$   takes a simpler form in the grand-canonical ensemble,
 where one sums over the rank $N$.\foot{The terminology ``grand-canonical ensemble'' originates from the Fermi gas approach to protected observables in the ABJM theory. It   has also   
 a clear thermodynamic interpretation in the M-theory bulk description below.}
 Defining the chemical potential $\mu$ conjugate to the rank $N$, the grand-canonical partition function is given by 
\begin{equation}\la{4j}
	\Xi(\mu,k) = \sum_{N=0}^\infty \rme^{\mu N} \rZ(N,k) = \sum_{n\in \mathbb Z}\rme^{J(\mu + 2\pi \rmi n)}\,,\quad \quad  J= J_{\rm pert}(\mu)  +  J_{\rm np}(\mu) \ , \quad   J_{\rm pert} =\tfrac{1}{3}  C\mu^3 + B \mu +A,
\end{equation}
and the Wilson loop expectation value takes the form
\begin{equation}
 \langle W(\mu,k)\rangle_{\text{grc}} = \frac{1}{\Xi(\mu,k)} \sum_{N=0}^\infty \rme^{\mu N}\, \rZ(N,k)\, \langle W(N,k)\rangle_{\text{can}} = \frac{1}{2\sin{2\pi\ov k} }e^{{2\ov k} \mu}+\OO(e^{-\mu}) \ .
 \label{5}
\end{equation}
Here  the $\OO(e^{-\mu})$  terms   are non-perturbative at large $\mu$, i.e. 
the expectation value is perturbatively 1-loop exact.\foot{See,
 for instance, \cite{Boldis:2025yll} and references therein for the discussion of such non-perturbative contributions.} 
To compare the leading large-$N$ behavior in the two ensembles we note that in the grand
 canonical ensemble $N$ is an expectation value 
 which at  the saddle point equals to 
\begin{equation}\la{5b}
	N = \partial_\mu J(\mu_\star,k) = C \mu_\star^2 + B\,.
\end{equation}
At leading order in large $N$, evaluating the grand-canonical result at the saddle $\mu_\star$ reproduces the same exponential and the same  sine prefactor as the canonical result.  The difference starts at order $N^{-1/2}$.

\subsubsection*{M2-brane  partition function }\label{s11}

The bulk observable dual to the WL expectation value discussed above is an M2-brane wrapping AdS$_2$ inside AdS$_4$ and the 11d Hopf circle of $S^7/\mathbb{Z}_k$. 
Its on-shell action  and 1-loop partition function  indeed  reproduce  the exponential   and the  sine  prefactor 
in \rf{3}
 \ci{Giombi:2023vzu}. 
 
 The aim of this paper is to compute the 2-loop correction starting from the M2 brane BST action \ci{Bergshoeff:1987cm} in its $\AdS_4 \times S^7$ form \ci{deWit:1998yu} (see also \ci{Gomis:2008jt}). While the semiclassical 1-loop correction in the 3d M2 brane effective action has no logarithmic UV divergences and is thus well defined (see, e.g., \ci{Duff:1987cs,Seibold:2024oyr,Beccaria:2023ujc,Giombi:2024itd,Beccaria:2025vdj,Beccaria:2025npl}), this theory is,  in general,  non-renormalizable starting at 2-loop order \ci{Beccaria:2025xry}. 
 Certain observables, such as partition functions of wrapped membranes in supersymmetric 11d backgrounds, may nevertheless remain free of logarithmic UV divergences \ci{Beccaria:2025ahf,Tseytlin:2026ctl}. 
 The same occurs in the present case of M2 brane on $\AdS_2 \times S^1 \subset \AdS_4 \times S^7/\ZZ_k$, with the  structure of the 2-loop   computation being similar to the one in the case of  M2 brane on $\AdS_3  \subset \AdS_7 \times S^4$ in \ci{Beccaria:2025ahf}.  


We will  find that the 2-loop  correction to the M2  brane partition function vanishes (i.e. $f_2=0$, see below). 
 In contrast to the $\AdS_3$ case in  \ci{Beccaria:2025ahf}   where the bosonic, mixed and fermionic 2-loop   contributions    each   had only power divergent parts   and thus were  separately regularized  to  zero,  here the  vanishing  will  be due to a   non-trivial   cancellation  between the bosonic and fermionic contributions and thus should be a reflection  of the underlying supersymmetry.


As discussed  below,  this result  disagrees  with the  expansion \rf{3} of the canonical ensemble expression for $\langle W \rangle$ 
 but agrees with \rf{5},  i.e.  is perfectly 
  consistent   with the proposal in \cite{Gautason:2025per,Gautason:2025plx} (see also \cite{Gautason:2025bft,vanMuiden:2026nsp,Bobev:2026gir}) that the semiclassical  M2  brane   computation in the 11d bulk theory 
  is to be compared directly to the grand-canonical ensemble in the dual boundary theory.

We use the convention in which $R$ is the radius of the round $S^7$ and  the $\AdS_4$ radius is $R/2$ so that the 11d   background  is 
%
\begin{align}
 \dd s_{11}^2&=R^2\big(\tfrac14\dd s^2_{\AdS_4}
 +\dd s^2_{S^7/\ZZ_k}\big) \ ,\qquad \qquad \dd s^2_{S^7/\ZZ_k}=\dd s^2(\CP^3)
 +(\dd\varphi'+A)^2, 
 \la{66}\\
& \qquad \varphi'\equiv \varphi'+\te {2\pi\ov k}\ , 
 \qquad \dd A=2J_{\CP^3} \ , \qquad  F_4=dC_3 = \tfrac{3\ii}{8} R^3 {\rm vol} (\AdS_4) \ . 
 \label{6}
\end{align}
In  Euclidean signature  the bosonic part  of the M2  brane action is\foot{Note that Euclidean continuation is applied to both target space and world-volume, so that $C_3$ here is imaginary and the WZ term in the M2 brane action is thus real.}
\begin{equation}
 S =T_2\int\dd^3\sigma\Big[  \sqrt{h}  +  \tfrac{\ii}{3!}\epsilon^{ijk}C_{MNP}
 \partial_iX^M\partial_jX^N\partial_kX^P+\ldots\Big],
 \qquad\qquad 
 T_2=\frac1{(2\pi)^2\lp^3}.
 \label{7}
\end{equation}
As in \ci{Sakaguchi:2010dg,Giombi:2023vzu}
we 
consider the classical M2 brane 
 solution  that wraps a unit-radius $\AdS_2\subset\AdS_4$ and the
Hopf circle  in \rf{6}.
We shall use the   static gauge    that  identifies the first two world-volume coordinates  $\s^1,\s^2$ 
 with $\AdS_2$   subspace of 
$\AdS_4$  and the 
 third  coordinate $\s^3$  with $2\pi$-periodic  $\varphi=k\varphi'$ in \rf{6}.  
 
 The M2  brane action  will depend on $k$ only via 
 the period   of  rescaled $\s^3$ coordinate 
 $ {1\ov k} \s^3$. To compute   quantum  corrections  one may start   directly  with the  $\AdS_4 \times S^7$ supercoset action \ci{deWit:1998yu}  and expand it  in fluctuations near the minimal surface. 
 The induced world-volume   metric   is then
 \be \la{8}
 ds^2 = h_{ij}d \s^i  d \s^j =\frac{1}{4} R^2  ds^2_{3}   \ , \qquad 
 ds^2_{3} = \dd s^2_{\AdS_2} + dy^2 \ , \ \ \ \ \  y= {2\ov k} \s^3, \ \ \   y\equiv y +  L \ , \ \ \   L={4\pi \ov k} \ , 
 \ee 
 where the  $\AdS_2$ subspace  has   unit radius so we rescaled the 3rd coordinate  by 2 defining $y= {2\ov k} \s^3$. Then $k$   will enter the quantum effective action  as a factor 
 in the  overall   volume  integral (that will factorize  as the  background is homogeneous)  and also non-trivially 
via   the momenta in the third direction ($p_n={2\pi n \ov L} = {1\ov 2} kn$).  The classical   value  of the 
action is then given by 
\be \la{9}
S_0 = T_2 V_3   = - {\pi^2 \ov k} \rT_2 \ , 
\qquad V_3 =   ({\te  {1\ov 2 }} R)^3   (-2\pi) L \ , \qquad\qquad   \rT_2 = R^3 T_2 = {R^3 \ov (2\pi)^2 \lp^3} \ . 
\ee
The    M2   brane path integral expanded    near the $\AdS_2 \times S^1$ 
minimal surface   is expected to reproduce the WL   expectation   value in the dual gauge theory 
\al{ \la{10}
&Z
=  \langle W\rangle= e^{-F} \ ,  \ \ \ \ \ \ \ \ \ 
 F= -\log Z =  \rT_2 f_0(k) +  f_1 (k) +   \rT_2^{-1} f_2(k) + \OO(\rT_2^{-2}) \ ,  \\
  &\qquad \qquad  \te   f_0 = - {\pi^2 \ov k}\ , \ \ \ \qquad \ \   \qquad    f_1 =-\log Z_1 = \log \big(2\sin{2\pi\ov k} \big) \ , \la{11}
}
where $f_1$  is the result of the  1-loop computation in \ci{Giombi:2023vzu}. 

%

\subsubsection*{Comparison to $\langle W\rangle$  in the two ensembles}

To   compare the M2 brane  expansion \rf{10} with the ABJM localization expressions for the WL in \rf{3} or \rf{5} 
one must   specify  which boundary variable is held fixed.

\textit{At fixed $N$}, one must first relate the membrane tension $\rT_2$ to the rank $N$. At the classical 11d supergravity level, the scale $R$ in \rf{66} is determined by the
 4-form electric charge, or, equivalently, by the number of M2 branes sourcing the background \ci{Duff:1990xz},
\begin{equation}\la{12} 
 R^6=32\pi^2N k\,\lp^6\ , \qquad {\rm i.e.} \qquad \bar \rT_2 = R^3 T_2 = {R^3 \ov (2\pi)^2 \lp^3} = {1\ov \pi} \sqrt{2 N k}\ .
\ee
Using this relation, the fixed-$N$ localization result \rf{3} may  be written  as
\be\la{4}
 \langle W(N,k)\rangle_{\rm can}
 =\frac{1}{2\sin{2\pi\ov k}}\,e^{{\pi^2\ov k}\bar\rT_2}
 \Big[1-\frac{1}{\bar\rT_2} f_{2,\rm can}(k)
 +\OO(\bar\rT_2^{-2})\Big]\,,\qquad f_{2,\rm can}(k)=\frac{k^2+32}{24k}\ .
\ee
Quantum M-theory corrections may, however, renormalize the tension--rank relation. Following \ci{Bergman:2009zh}, one is to  replace $N$ in the definition of the radius by $\NN=N-\frac{k^2-1}{24k}$ :\foot{The M2 brane   perturbation theory  is defined by the   tension   that  appears as the coefficient in the action in the path integral. We  will assume   that  it is 
 $\rT_2$ as in \rf{9}, i.e. will  ignore the notational clash between  \rf{9}  and \rf{12}.}   
%
\begin{equation}
 \RR^6=32\pi^2\NN\, k\,\ep^6\ , \qquad
 \rT_2 = {\RR^3\ov (2\pi)^2 \lp^3}= {1\ov \pi} \sqrt{2\NN k}\ , \qquad
 \NN=N-h(k)\ , \qquad h(k)=\frac{k^2-1}{24k}\ .
 \label{13}
\end{equation}
If $F$ in \rf{10} is written in terms of $\rT_2$ and then re-expanded in terms of the classical tension $\bar\rT_2$ in \rf{12}, the coefficients $f_0$ and $f_1$ are unchanged, while
\be
 f_2 \ \longrightarrow\ f_2-\te {1\ov\pi^2}k\,h(k)f_0=f_2+h(k)\ .
 \la{14}
\ee
Comparison with the fixed-$N$ coefficient $f_2^{\rm (can)}$ in \rf{4} would therefore require the local M2 brane calculation to give
\be
 f_2\overset{?}{=}{11\ov8k}\ .
 \la{15}
\ee
Instead, our explicit computation  below will give $f_2=0$, and there is no natural redefinition of $N$ that  would convert it  into the canonical  ensemble   coefficient in \rf{3}. 

\textit{At fixed $\mu$}, we instead are to relate the bulk and boundary parameters through the boundary value of the three-form potential $C_3$, rather than the dual flux through $S^7/\ZZ_k$ that fixes $N$.  Fixing  the  gauge by requiring  regularity in the interior of Euclidean $\AdS_4$, one finds in the conventions of \rf{6}\footnote{The chemical potential is defined up to large gauge transformations. The final answer thus has to incorporate a sum over the large gauge images, which is the bulk incarnation of the sum over $n$ in \rf{4j}. We assume  a semiclassical regime of large $\mu$ where those large gauge images are non-perturbatively negligible.\label{FN 4}} 
\be\la{5a}
 \mu=-\ii T_2\Big[\int_{\partial\AdS_4}C_3\Big]_{\rm ren}
 ={R^3\ov8\lp^3}\ . 
\ee
The corresponding dimensionless membrane tension and 1-loop partition function are\footnote{Taking into account the large gauge transformations, as mentioned in footnote \ref{FN 4}, the membrane partition function equals
\begin{equation}
	Z_{\text{M2}} (\mu) = \frac{\sum_{n\in \mathbb{Z}} \rme^{J(\mu + 2\pi \rmi n)}
	 Z(\mu + 2\pi \rmi n)}{\sum_{n\in \mathbb{Z}} \rme^{J(\mu + 2\pi \rmi n)}}\,.\no
\end{equation}}
\be\la{5a2}
 \rT_2^{(\mu)}=R^3T_2={2\mu\ov\pi^2}\ , \qquad\qquad 
 Z(\mu,k) 
= \frac{1}{2\sin{\frac{2\pi}{k}}} \rme^{\frac{2\mu }{k}}\,.
\ee
Once again,  the classical action reproduces the exponential $e^{2\mu/k}$ in \rf{5}, while the 1-loop determinant gives its sine prefactor. The result $f_2=0$ then agrees with the absence of further perturbative corrections in the grand-canonical ensemble result in   \rf{5}. 

The vanishing of the 2-loop M2 brane correction thus provides   strong evidence in support of the proposal of \ci{Gautason:2025per,Gautason:2025plx} that the semiclassical M2 brane partition function
 corresponds, in the dual gauge theory,  to the grand-canonical ensemble description at fixed $\mu$, rather than the canonical ensemble description  at fixed $N$.


\subsubsection*{String theory limit}

The type IIA string theory limit provides a useful consistency check of the M2 brane result. In the canonical ensemble,  in this limit, $N$ and $k$ are taken large with $\lambda=N/k$ fixed, and the Wilson loop is described by an $\AdS_2$ fundamental string worldsheet in $\AdS_4\times\CP^3$. The localization result in \rf{1},\rf{3} becomes
 \be\la{811}
\langle W \rangle_{\rm planar} = \frac{N}{4\pi\l}e^{\pi\sqrt{2(\l- {1\ov 24})}}  +  \mc O(N^{0}) 
 = \frac{N}{4\pi\l}e^{\pi\sqrt{2\l}}  \,   \Big(1-\frac{\pi}{24\sqrt 2}\frac{1}{\sql}+\dots    \Big)+\mc O(N^{0}) \ . 
\ee
The grand-canonical parametrization gives a direct interpretation of the shifted coupling in \rf{811}. At the saddle $N=\langle N\rangle_\mu$, with $\lambda=\langle N\rangle_\mu/k$, define
\be\la{812}
 \l_\mu\equiv {\langle N\rangle_\mu-B(k)\ov k}
 =\l-{1\ov24}-{1\ov3k^2}\ ,\qquad\qquad
 {2\mu\ov k}=\pi\sqrt{2\l_\mu}
 \ .
\ee
The  $k^{-2}$ term in $\l_\mu$ vanishes in the strict type IIA limit, so that $\l_\mu\to\l-\frac1{24}$. Thus the shifted coupling appearing in the  string tension is already encoded in the saddle-point relation between the chemical potential and the mean rank. 

Correspondingly, the natural definition of the dual string tension, obtained from the large-$k$ limit of \rf{13}, is
   \iffa 
   \foot{An  indication that this is the correct relation
for string tension  comes from the  the expression for the  magnon dispersion relation  and  from the cusp anomaly 
at 2-loops \cite{Bianchi:2014ada}: the  cusp anomaly   takes a   natural form in terms of effective tension  given by 
$T= \frac{1}{\sqrt 2}\sqrt{\l-\frac{1}{24}}-\frac{\log 2}{2\pi}$  (the $\log 2$  term is not   relevant  in WL  case as it  leads just 
to an extra overall  rescaling).
 This is also   suggested by   the structure of localization matrix model that  implies  
   that  the combination $\l- \frac{1}{24} +  \mc O(1/N)$     plays  a special role. 
This follows  from the Fermi gas representation   $\exp(-\mu N  + J)$, 
 $J= \frac{1}{3} C \mu^3  + B \mu + A$, \ $B= \frac{k}{24}+\frac{1}{3k}$  (for a review see, e.g.,  \cite{Beccaria:2023ujc}).
This suggests an  effective shift $N\to N-B = N-\frac{k}{24}-\frac{1}{3k}$. 
 Also,  the argument in \cite{Bergman:2009zh} suggests 
that AdS  radius in IIA limit should be  given by 
$L^4 =  \l - \frac{1}{24}   + \frac{1}{24}1/k^2$    which translates into  the  expression for the 
effective string  action  at  large $N$   as  $T = 1/\sqrt 2 \sqrt{\l - 1/24}$.}
\fi 
\be
\la{821}
\rT_1=  \tfrac{1}{\sqrt 2}\sqrt{\l- \te {1\ov24}} = \bar \rT_1 \Big(1 - \frac{1}{ 48 \l } + ... \Big), \qquad  \ \  \bar \rT_1\equiv {R_{\AdS}^2\ov 2\pi  \ls^2} =   \frac{\sqrt \l}{\sqrt 2} \ ,  \ \ \ \ \  \ \ \  R_{\AdS} =\ha R \ . 
\ee
Introducing the string coupling $g_{\rm s}= \sqrt \pi ( 2 \l)^{5/4}/N$ \ci{Aharony:2008ug}, the planar result \rf{811} may then be written as \ci{Giombi:2020mhz}
\be\la{831}
Z_s= \langle W \rangle_{\rm planar} = Z_{1s} 
\ e^{2\pi \rT_1  }  +  \OO\big(g^0_{\rm s}\big) \ , \qquad \qquad  Z_{1s}=\frac{\sqrt{\bT_1} }{\sqrt{2 \pi} \ g_{\rm s} }
\ . 
\ee
The prefactor $Z_{1s}$  has a useful interpretation from the M2 brane perspective. As discussed in \ci{Giombi:2020mhz}, a direct 1-loop GS string computation of the $\AdS_2$ partition function in $\AdS_4\times\CP^3$ \ci{Kim:2012tr} requires a particular measure or counterterm contribution that cancels a topological UV divergence  (the same issue arises in  the $\AdS_5\times S^5$ string case).

It was observed in \ci{Giombi:2023vzu,Seibold:2024oyr} that this contribution is generated automatically when the string path integral is defined as the $k\to\infty$ limit of the M2 brane path integral: it arises from the regularized sum over the non-zero KK levels. Indeed, the 1-loop term in \rf{831} is the leading large-$k$ term in the M2 brane 1-loop factor \ci{Giombi:2020mhz}, i.e. $Z_1= (2\sin { 2 \pi \ov k})^{-1} = {k\ov 4\pi} + ...$, which matches $Z_{1s}$ in \rf{831}.
   
The corresponding string free energy on the disk takes the form (cf. \rf{3})
\be 
F_s= -\log Z_s =  - 2 \pi \rT_1 - \log [  \tfrac{\sqrt{\rT_1} }{\sqrt{2 \pi} \ g_{\rm s} }  ]      - \tfrac{1 }{  192}  \rT_1^{-2} 
 + \OO(\rT_1^{-3}) \ . \la{841}
\ee
Note that  there is no 2-loop $\rT_1^{-1}$ correction. To compare this 
 with the M2 brane expansion, recall that double dimensional reduction relates the naive and renormalized string and membrane tensions according to (cf. \rf{13}, \rf{821})
\be 
\bar \rT_1 = {1\ov 4} {2\pi \ov k} \bar \rT_2   \ , \qquad \qquad   
 \rT_1 = {\pi\ov 2} \big({1\ov k}  \rT_2\big) \Big|_{k\to \infty}\ .  \la{120} \ee
It follows that the 2-loop coefficient $f_2$ in the M2 brane expansion \rf{10} is related to its string counterpart in \rf{841}, defined by $F_s =\rT_1  f_{0s} + f_{1s} + \rT_1^{-1} f_{2s} + ...$, as
$f_{2s} = {\pi\ov 2} \big({1\ov k}  f_2\big) \big|_{k\to \infty}$. Hence any $f_2$ that grows at large $k$ as $k^r$ with $r<1$ gives a vanishing string 2-loop correction, in agreement with \rf{811},\rf{841}. This includes both our result $f_2=0$ and the fixed-$N$ candidate in \rf{15}. The string theory limit is therefore a consistency check of the calculation, but by itself does not distinguish between the two ensemble prescriptions.
      
  \iffa
According to \cite{Gautason:2025plx}    starting with the Fermi  gas expression   for 
$\langle W\rangle$  one should not perform  the integral over $\mu$  variable "dual" to $N$   so that 
 $\langle W\rangle = \frac{1}{2\sin\frac{2\pi}{k}}e^{2\mu/k}+\mc O(e^{-\mu})$
where $\mu$    should  be effectively replaced by  $\frac{\pi}{\sqrt 2}\sqrt{Nk}$.
 The exponential factor 
is then $e^{\pi\sqrt{2\l}}$  without  shift  in $\l$.\foot{From a broader perspective of comparison of ABJM results with dual string or M2 brane theory 
it is not  unclear how to reconcile this  prescription 
with  non-zero 2-loop  string correction to cusp anomaly  \cite{Bianchi:2014ada} and 
with the expected  shift in $N$ or AdS$_4$   radius in \cite{Bergman:2009zh}.}
\fi

\subsubsection*{Two-loop M2 brane  contribution}\label{s12}

Expanding   M2  brane action  near  the $\AdS_2 \times S^1$ minimal surface in the static gauge 
we find that 
 like  for the  M2  brane in static gauge in  flat  target space 
and also for  AdS$_3$   M2  brane in $\AdS_7 \times S^4$   
  \ci{Seibold:2024oyr,Beccaria:2025xry,Beccaria:2025ahf}
 there  is 
a  natural $\kappa$-symmetry gauge  in which  there are no cubic couplings in   the action,  
\be \la{160}
S= S_0 +\int d^3\s\, \sqrt g \, \L + ... \ , \ee
  where $S_0$ is the classical value \rf{9}   and 
the   Lagrangian $\L$ for (rescaled)  8+8 bosonic and  fermionic fluctuations which are 3d fields   is given by 
\al{ \la{16}
&\L= \L_2 + 8\rT_2^{-1} \L_4+\OO(\rT_2^{-2})
 \ , \qquad\qquad  \L_2 = \L_{2 b} + \L_{2f} \ , \qquad \qquad  \L_4 = \L_{4 b} + \L_{2b,2f} + \L_{4f} \ , \\
&\L_{2b}  = |\nabla x|^2+|\nabla_y^{(-\frac32)}x|^2-\tfrac14|x|^2 +
 |\nabla w^a|^2+|\nabla_y^{(\frac12)}w^a|^2-\tfrac14|w^a|^2
   \ ,   \qquad \ \ \  a=1,2,3 \ , \la{17} \\
&\cL_{2f}
 =\bar\chi\big(\slashed\nabla+\gamma^3\partial_y\big)\chi
 +  \bar\psi^a\big[\slashed\nabla+\gamma^3\nabla_y^{(-1)}\big]\psi^a \ , \qquad \qquad \nabla_y^{(q)}\equiv \partial_y+\ii q\ .
 \la{18}
}
 $\L_2$ corresponds to the metric  $ds^2_3$ in \rf{8}  
 ($\nabla$ is the $\AdS_2$ covariant derivative).  
The bosonic fluctuations are one complex field $x$ from the transverse
$\AdS_4$ directions and three complex fields $w^a$  from
$\CP^3$. The  gauge-fixed  fermions in the BST action   can be organized as one complex
3d Dirac  fermion $\chi$ and three complex Dirac  fermions  $\psi^a$.
Expanding in  Fourier modes in $y\equiv y + {4\pi \ov k}$  one 
finds  8+8   towers of  $\AdS_2$ fields 
 with masses  \ci{Sakaguchi:2010dg,Giombi:2023vzu}
\al{
m^2_{x,n} = (p_n-1)(p_n-2) ,  &\ \   \ \ \  
m^2_{w,n} = p_n (p_n+1) , \ \ \ \ \  m_{\chi,n} = p_n \ , \ \ \ \ \ 
m_{\psi,n} = p_n -1 , \la{17b}\\
&p_n=\te {2\pi n \ov L}= {\ha k \, n  }  \ , \ \qquad  \  k > 2 \ , \ \ \ \ \ \ \ \  \ \ \ n=0,  \pm 1,\pm 2 ... \ . 
\la{18b}}
The quartic interaction term in \rf{16}  has  the  following symbolic structure (here $X= ( x, w^a)$ and $\theta=(\chi, \psi^a)$)
 \ba  
   \L_{4} =  \Big[(\del X)^4  + X^2 (\del X)^2 + X^4 \Big] +   
    \big[ (\del X)^2  + X^2\big] (\theta \nabla \theta + \theta^2 )  + 
     \Big[ \theta  \theta \theta  \nabla \theta   +   \theta\nabla  \theta\theta  \nabla \theta \Big]    . \la{19}
   \ea
   As a result, 
the 2-loop diagrams  that  contribute to the  free energy 
are just the double-bubble  ``OO'' ones, 
  given by products  of  (derivatives of)  two  bosonic, one bosonic  and one fermionic    and     two fermionic  propagators in  $\AdS_2 \times S^1$  at coinciding points.  
  Then $f_2$ in \rf{10} is simply proportional 
  to $\langle \L_4 \rangle$.

    The corresponding contribution to the  free   energy in \rf{10}  is 
  \al{
F=S_0-\log Z_1+{8}{\rT_2}^{-1} \,\langle S_4\rangle
 +\OO(\rT_2^{-2})\ , \qquad \qquad  S_4=\int_{\Sigma_3} d^3\s\,\sqrt g\;\cL_4\ .
 \la{29}
}
Since the background is homogeneous, $\langle\cL_4\rangle$ is a constant
and $\langle S_4\rangle=\bar V_3\,\langle\cL_4\rangle$, where $\bar V_3  $ is the
renormalized volume of  the 3d metric $ds^2_3$ in \rf{8} 
\be
 \bar V_3=\int_{\Sigma_3} d^3\s\,\sqrt g = L\,\vol(\AdS_2)=-\tfrac{8\pi^2}{k}\ .
 \la{30}
\ee
As a result,  $f_2$ in \rf{10}   is given by\foot{Note that the classical value of the induced metric in \rf{8} is $h^{(0)}_{ij}= {1\ov 4} g_{ij}$ so that  $\sqrt h^{(0)}= {1\ov 8} \sqrt g $. The fluctuation fields are rescaled by factor of $\sqrt{ 8 \rT_2^{-1}}$. }
\be
 f_2=8\,\bar V_3\,\langle\cL_4\rangle
 =-\tfrac{64\pi^2}{k}\,\langle\cL_4\rangle\ .
 \la{31}
\ee
Since  in 3d the  propagators have no log divergences,  any UV   divergences in $f_2$
can therefore only be power divergences. The latter   can   be regulated away  by  an analytic regularization such as 
the $\zeta$-function  one that we will use below. 
If one  first expands in  Fourier modes on $S^1$   getting towers of fluctuations on $\AdS_2$ 
 then the log UV divergences  present   in $\AdS_2$   propagators will  go away after  summation over the mode number $n$ (using that $\sum_{n=-\infty}^\infty 1 = 1 + 2 \zeta(0) =0$). 

 Let us  introduce the notation for the $\AdS_2$  propagator constants for massive bosons and fermions
 ($\Lambda$ is UV cutoff in  $\AdS_2$ theory and $\uppsi(x) = \Gamma'(x)/\Gamma(x)$)
 \al{
& \G_b(m)= (-\nabla^2 + m^2)^{-1}\Big|_{\s\to \s'}   =\tfrac{1}{2\pi}\Big[ \log\Lambda-   \uppsi(\Delta)\Big] \ ,\ \qquad  \qquad \ \ \
\qquad \Delta= \ha + \sqrt{ m^2 + \tfrac{1}{4}} \ , \la{20} \\
& \G_f(m) = \ii \gamma^3 (\slashed\nabla+\ii\gamma^3 m)^{-1}\Big|_{\s\to \s'}   = \tfrac{1}{2\pi} m \Big[
  \log\Lambda-   \uppsi(|m|)-\frac{1}{2|m|}\Big]\ , \qquad    \Delta= |m| +\ha \ .\la{020}
 }
 From  \rf{17b} we have ($p_n =\ha k n$)
 \al{\Delta_{x,n}=\tfrac12+\big|p_n-\tfrac32\big|, \qquad  
 \Delta_{w,n}=\tfrac12+\big|p_n+\tfrac12\big|, \qquad 
  \Delta_{\chi,n}= 
  |p_n| + \ha , \qquad 
 \Delta_{\psi,n}= |p_n -1| +\ha  \ .  \la{21}
  }
 As we  will  find below, 
   explicit evaluation of the quartic terms in the M2 brane
action shows that, 
     remarkably,   the 
  expectation values  $ \langle\cL_{4b}\rangle $, $ \langle\cL_{2b,2f}\rangle $
   and $ \langle\cL_{4f}\rangle $  expressed in terms of  products of  two bosonic and fermionic  3d Green's functions 
    combine  together   to  form a  ``perfect square''  combination  
   \al{
&\qquad \qquad  \langle\cL_4\rangle=- \tfrac{3}{8} \cX^2 \ ,  \qquad 
\qquad \cX=\cX_b-\cX_f=\tfrac{k}{4\pi}\sum^\infty_{n=-\infty}\X_n\ ,
 \qquad\qquad   f_2 =\tfrac{24\pi^2}{k}\cX^2 \ , 
 \label{22}\\
& \X_n=\Big[{2p_n(p_n-1)\G_b( m_{x,n}  )}
 +{6p_n(p_n+1)\G_b(  m_{w,n}   )}\Big] 
 - \Big[{2(p_n+1)\G_f (m_{\chi,n} )} +{6p_n\G_f (m_{\psi,n} )}\Big]
  \ . 
 \label{23}
}
Using \rf{20},\rf{21}  and that $p_n=\ha k n $   we find 
 that all  $\log \Lambda$ terms 
and  all transcendental constants   cancel  for each $n$  in the combination of  levels $n$  and $-n$
and thus we get (for $k>2$) 
\be 
\X_n + \X_{-n} =\te - {1\ov \pi} \ , \ \ \ \ \ \ \ \ \ \   n\ne 0 \ ;  \qquad \ \ \ \ \ 
\X_0=    - 2   \G_f (0)  
   \ .                 \la{24}  \ee 
The non-zero part of the ``level''  $n=0$  term 
$\X_0$ is given  just by the contribution of the   massless   fermion $\chi$.
The  definition of   the  value   $\G_f(0)$ of the massless fermion  Green's function at coinciding   points in $\AdS_2$ 
 is known to be  subtle \ci{David:2023btq} (see also \ci{Gripaios:2008rg}  and below).

If we  first  regularize the sum over $n$ in \rf{22} using Riemann   $\zeta$-function  keeping $G_f(0)$ 
 general we get 
\be \la{244}
\cX=\tfrac{k}{4\pi}\Big[ \X_0 -  \tfrac{1}{ \pi}  \sum^\infty_{n=1} 1 \Big]
=- \tfrac{k}{2\pi}\big[  \G_f (0)  -  \tfrac{1}{4\pi} \big]\ , 
\qquad \qquad \qquad 
f_2 =  6  k \big[   \G_f (0)  - \tfrac{1}{4\pi}\big]^2  \ . \ee
As $  \G_f (0)$ is just a $k$-independent   constant  this expression  for $f_2$   has 
different $k$-dependence from the value  \rf{15}  required by matching to the canonical-ensemble  localization result.

\vthree In fact, the correct value   of  $\G_f (0)$   is  fixed  by the  supersymmetric  choice of the  boundary 
 domain of the massless Dirac operator, which  can be  determined  directly at $m=0$ (see section \ref{X-result}); 
 it agrees  with  the  massless  limit  in \rf{020}  taken  as $m \to 0^+$ (see  \ci{David:2023btq}). This gives (using that $
\uppsi(a\to0^+)=-\frac{1}{a}-\gamma_{\rm E}+\OO(a)$) 
\be \la{25} \te 
\displaystyle  \lim_{m\to 0^\pm } \G_f (m) \equiv \G_f(0^\pm) = \pm\tfrac{1}{4\pi}\ , \qquad \qquad 
 \G_f(0^+ ) = \tfrac{1}{4\pi}\ . 
\ee
The   choice of    $ \G_f(0^+ ) $  is  dictated by  the  supersymmetry-consistent   choice of 
 boundary conditions as  in the present case the  massless  bosons  $w^a_{0}$  satisfy the 
 Dirichlet b.c. (the  minimal surface   is localized in $\CP^3$). 
 As a result, we get from \rf{244}
 \be \la{26}
 \cX=0  \ ,\ \  \qquad {\rm i.e.} \ \ \ \qquad f_2=0 \ , \ee
  matching the  localization   prediction  in  the grand-canonical ensemble. 
 


\

The structure of the rest of the paper is as follows. 
Section~\ref{quadratic} presents the quadratic 3d theory  and mass spectrum. 
 Section~\ref{quartic-square}  discusses  expectation values of quartic terms in the action.
Section~\ref{X-result} evaluates the combination  $\cX$, isolates the
zero-level  ambiguity and fixes it  in a   way  consistent   with supersymmetry. 
  Section~\ref{discussion}  contains  some  concluding comments.
   There  are several Appendices  containing technical details of the derivations
   of  the main results.

\section{Quadratic Lagrangian}
\label{quadratic}


Our starting point   is the M2 brane action in  $\AdS_4\times S^7$. We  use the
supercoset action of \ci{deWit:1998yu} (see also  \ci{Claus:1998fh,Gomis:2008jt,Sorokin:2010wn,Tsimpis:2004gq,Wulff:2013kga})  and expand it
near the $\AdS_2\times S^1$ solution
of section \ref{s1},  using  the static gauge and  the
$\kappa$-symmetry gauge specified in Appendix~\ref{apa}. 
 \iffa 
 \foot{To compute the   string  or M2 brane partition function expanded  near  AdS$_2$  or AdS$_2\times S^1 $ 
minimal surface we will need to start with the corresponding GS  or BST  action. 
The type IIA  GS string   action in AdS$_{4}\times \CP^{3}$ expanded   in the static gauge will    have a similar form 
to the \adss action. 
The  supercoset construction  of the action was given in \ci{Arutyunov:2008if,Stefanski:2008ik}. It can also be   obtained by  double dimensional reduction of the action for M2 brane in AdS$_{4}\times S^7$  given in 
\ci{deWit:1998yu,Claus:1998fh} (see \ci{Uvarov:2009hf}; this  action in l.c. gauge was used in  2-loop  cusp anomaly computation 
 \cite{Bianchi:2014ada}). 
 A better starting point may be the action in \ci{Gomis:2008jt,Sorokin:2010wn} that has a similar form to the \adss 
   action.
   The quadratic in fermions part of the action was  discussed in \ci{Kim:2012tu}. To get the quartic fermionic term
   one may  use the general form of  expansion in $\theta$  of the  GS action in a IIA  supergravity background 
     given in \ci{Wulff:2013kga}   that   is related  by reduction 
     to the   quartic terms 
     for general 11d background in \ci{Tsimpis:2004gq}.
} 
\fi 
As already  mentioned above, 
 the
orbifolding by $\ZZ_k$ acts only globally: it enters through the
identification $y\equiv y+L$, \ $L=\frac{4\pi}{k}$ in the induced metric
\rf{8}  and the volume factor. 
  The resulting  3d  fluctuation   action  is covariant   with  respect to the unit-radius product metric
$ds^2_3$ in \rf{8}.  

The quadratic Lagrangian of the $8+8$ fluctuation fields was  already given in
\rf{17},\rf{18}:  there is one complex scalar $x$ from the two $\AdS_4$ directions
transverse to $\AdS_2$, three complex scalars $w^a$ ($a=1,2,3$) from
$\CP^3$, one complex 3d Dirac fermion $\chi$ and a triplet
$\psi^a$. They have  effective   $S^1$  charges  given by 
$q_x=-\frac32$, $q_w=+\frac12$, $q_\chi=0$, $q_\psi=-1$ \ ($\nabla_y^{(q)}=\partial_y+\ii q$),
see Appendix~\ref{apb}  for details.\foot{Note 
 that  the   scalar   mass terms in \rf{17}  correspond to the 3d  conformal coupling 
$\frac18 R^{(3)}=-\frac14$  where 
$R^{(3)}=-2$  is the curvature of $ds^2_3$ in \rf{8}.}

Expanding in  Fourier  modes  in $y\equiv y + L$, i.e. 
\be \Phi(y)=\tfrac1{\sqrt L}\sum_{n=-\infty}^\infty\Phi_n\,e^{\ii p_ny} \ , \qquad \qquad 
\te p_n = {2\pi n  \ov L}= {\ha k n } \ , \qquad \ \  L= {4\pi\ov k}\ ,  \la{32} \ee
one finds that for $k >2$  the  towers of $\AdS_2$ fields  have the following
quadratic Lagrangians \ci{Sakaguchi:2010dg,Giombi:2023vzu}\foot{Here the fermionic KK modes are complex  2d Dirac  spinors. The bar includes both charge conjugation
and reversal of the Fourier level. More precisely,
if $\chi_n=\Pi_{\T}\theta_n$, where $\Pi_{\T}$ are the projectors on the 11d Majorana spinor defined below, then $\bar\chi_n$ in   \rf{34}    is a shorthand for $\bar\theta_{-n}\Pi_{\T}=(\theta_{\T^c,-n})^TC$,  where
$C^{-1}\Pi_{\T}^{\,T}C=\Pi_{\T^c}$.
Thus the original 11d Majorana modes are paired as
$(\T,n)\leftrightarrow(\T^c,-n)$; the same convention applies to
$\psi_n^a$. See \rf{131} for the action of charge conjugation on the
projectors, \rf{141} for the Fourier-mode convention, and \rf{156} for
the explicit relation between the barred and unbarred KK modes.}
\al{
& \cL^{(n)}_{2b}=|\nabla x_n|^2+ m^2_{x,n} \,|x_n|^2
 + |\nabla w^a_n|^2+ m^2_{w,n}\,|w^a_n|^2 \ , \la{33}\\
 \qquad 
& \cL^{(n)}_{2f}=\bar\chi_n\big(\slashed\nabla+\ii\gamma^3 m_{\chi,n}\big)\chi_n
 + \bar\psi^a_n\big(\slashed\nabla+\ii\gamma^3 m_{\psi,n} \big)\psi^a_n\ ,
 \la{34}
}
 with the masses in \rf{17b}    summarized again  in the table below\foot{
Here $\ii\gamma^3$ is the
two-dimensional chirality matrix, and the fermion mass parameter is defined as the coefficient of $\ii\gamma^3$. In~the notation of Appendix \ref{apb}, $\gamma^3$ is the restriction of $-\ii \Gamma_1\Gamma_2$ to each $\kappa$-gauge fixed two-dimensional spinor block.}
\begin{center}
\begin{tabular}{@{}lccc@{}}
\toprule
field & multiplicity & $\AdS_2$ mass & value at $n=0$ \\
\midrule
$x$ & 1 complex & $m^2_{x,n}=(p_n-1)(p_n-2)$ & $2$ \\
$w^a$ & 3 complex & $m^2_{w,n}=p_n(p_n+1)$ & $0$ \\
$\chi$ & 1 complex Dirac & $m_{\chi,n}=p_n$ & $0$ \\
$\psi^a$ & 3 complex Dirac & $m_{\psi,n}=p_n-1$ & $-1$ \\
\bottomrule
\end{tabular}
\end{center}
The masses obey the two standard $\AdS_2$ supersymmetry relations
\be
 m^2_{x}=m_\psi(m_\psi-1)\ ,
 \qquad\qquad 
 m^2_{w}=m_\chi(m_\chi+1)\ .
 \la{35}
\ee
Keeping track of the sign of the Dirac mass, the  masses of the fermion family  $\{\vT\}$ at
fixed level $n$ (including the charge conjugates and multiplicities) are
\be\la{37}
 m_{\vT,n}=\big\{\,(p_n-1)\ (\times3),\ \ ( -p_n-1)\ (\times3),\ \ p_n,\ \ -p_n\,\big\} .
\ee
Charge conjugation does not act as $m\to-m$ at fixed Fourier level; instead, we have 
\be
 m_{\vT^c,-n}=m_{\vT,n}\ ,
 \la{39}
\ee
where $\vT^c$ indicates reversal of  the charge sign. 

The world-volume supersymmetry of the M2 brane arises from the preserved target-space Killing spinors
\cite{Kallosh:1997sw,Kallosh:1997ky}. While the physical fermions
satisfy \eqref{119}, the BPS condition selects the target space Killing spinors through
\begin{align}
    \Gamma_{\kappa}\varepsilon=\varepsilon \ .
\end{align}
To write down a world-volume supersymmetry transformation explicitly,
it is convenient to choose a particular complexified preserved
Killing spinor satisfying, in the notation of Appendix~\ref{apb},\footnote{Note that the
 mutually commuting operators $\ss_r$ commute with the pulled-back Killing-spinor derivative, so the Killing spinors can be chosen to have definite $\ss_r$ eigenvalues.
 Note that  the sign relation among the $\ss_r$ eigenvalues is opposite to that for the physical fermion blocks in Appendix B, since the preserved Killing spinor obeys $\Gamma_\kappa\varepsilon=+\varepsilon$ whereas the gauge-fixed physical fermions obey $\Gamma_\kappa\theta=-\theta$.}
\begin{align}
    \ss_0\varepsilon=\ss_1\varepsilon=\ss_2\varepsilon=-\varepsilon
    \ , \qquad
    \ss_3\varepsilon=\varepsilon \ ,
    \label{sk1}
\end{align}
so that the pulled-back Killing-spinor equations \eqref{124} reduce to
\begin{align}
    D_{\alpha}^{(0)}\varepsilon
    =\nabla_{\alpha}\varepsilon
    +\tfrac{\ii}{2}\rho_{\alpha}\gamma^3\varepsilon=0
    \ , \qquad
    D_3^{(0)}\varepsilon=\partial_3\varepsilon=0 \ .
    \label{sk2}
\end{align}
With this choice of the supersymmetry parameter, the world-volume
supersymmetry can be written as
\begin{align}
    \delta_{\varepsilon}\bar{\Phi}
    &=\ii\bar{\Psi}\gamma^3\varepsilon \ ,
    \qquad
    \delta_{\varepsilon}\Psi
    =\big[\ii(\slashed{\nabla}\Phi)\gamma^3
    -(P+q_{\Psi})\Phi\big]\varepsilon \ ,
    \qquad
    \delta_{\varepsilon}\Phi
    =\delta_{\varepsilon}\bar{\Psi}=0 \ ,
    \label{sk3}
\end{align}
where $P\equiv-\ii\partial_y$ and $q_{\Psi}$ is the effective $S^1$
charge of $\Psi$, with $q_{\psi^c}=+1$ and $q_{\chi^c}=0$. The physical fluctuations organize into multiplet pairs
$(\Phi,\Psi)$ as\footnote{The pairings between different fields can be seen from the bosonic transformations $\delta_{\varepsilon}X^A\sim\bar{\theta}\Gamma^A\varepsilon$,
which can be further projected as in Appendix~\ref{apb}.
The resulting supersymmetry transformations for each pair
$(\Phi,\Psi)$ and for each KK mode on $\AdS_2\times S^1$ are
essentially those considered in \cite{Sakai:1984vm}, with an
additional $\gamma^3$ insertion.}
\begin{align}
    (x,\psi_3) \ , \qquad
    (\bar w^3,\chi^c) \ , \qquad
    (w^1,\psi_2^c) \ , \qquad
    (w^2,-\psi_1^c) \ .
    \label{sk4}
\end{align}
The supersymmetry transformations \eqref{sk3} act on each $S^1$ Fourier mode separately and leave the quadratic bulk action invariant up to boundary terms. Since $\AdS_2$ has a boundary, 
one has to specify the boundary conditions compatible with the supersymmetry transformation \eqref{sk3}.

In particular, if the scalar in the Fourier-mode pair
$(\Phi_n,\Psi_n)$ satisfies Dirichlet boundary conditions, the 
supersymmetry requires
\begin{align}
    \delta_{\varepsilon}\bar{\Phi}_n\big|_{\partial\AdS_2}
    =\ii\bar{\Psi}_n\gamma^3\varepsilon\big|_{\partial\AdS_2}
    =0 \ .
\end{align}
Using the leading Killing spinor radial behaviour determined by
\eqref{sk2} and taking its conjugate condition,
one obtains the necessary fermionic boundary condition as (see also \cite{Correa:2019rdk}) 
\begin{align}
    \lim_{z\to0}z^{-1/2}\bar{\Psi}_{n}\PPi_{-}=0 \ , \qquad \lim_{z\to0}z^{-1/2}\PPi_{+}\Psi_{-n}^{c}=0 \ , \qquad
    \PPi_{\pm}=\tfrac12\big(1\pm \ii\gamma^{\hat z}\gamma^3\big) \ .
    \label{sk5}
\end{align}
Here $z$ is the radial   $\AdS_2$ coordinate with the boundary being  at $z=0$,
and $\gamma^{\hat z}=z\rho_{z}$ is the gamma matrix along the unit
normal to $\partial\AdS_2$. The boundary condition \eqref{sk5} becomes 
 important when defining the fermionic propagators, as we discuss in section \ref{X-result}.

\section{Quartic Lagrangian  and its expectation value}
\label{quartic-square}

The quartic  Lagrangian  \rf{19} contains  a purely bosonic term, a mixed boson--fermion term and a
four-fermion term.  In this section we display these quartic 
 vertices and then  compute their  quantum expectation values. Details and  Wick  contraction  rules are 
collected in the Appendices. 

Here and in the Appendices it is more convenient to use the
$2\pi$-periodic coordinate $\s^3=\varphi$ instead of $y$, so that
$y={2\ov k}\s^3$ and $\del_3={2\ov k}\del_y$, cf. \rf{8}.
All $k$-dependence in the vertices below  can be absorbed back into
this coordinate redefinition.

\subsection{Bosonic sector}

Starting with \rf{66}   and expanding the induced 3d metric in fluctuations  as
\be 
h_{ij}=\tfrac14(g_{ij}+H_{ij})+h^{(4)}_{ij}+...\ , \la{311}\ee
 we get the quadratic
part as
\al{
 H_{ij}={}&\partial_ix\,\partial_j\bar x
 +\partial_jx\,\partial_i\bar x
 +\partial_iw^a\partial_j\bar w^a
 +\partial_jw^a\partial_i\bar w^a
 +2|x|^2P_{ij}
 +\tfrac4{k^2}\big(\delta_i^3J_j+\delta_j^3J_i\big)\ .
 \la{46}
}
Here $P_{\alpha\beta}=g_{\alpha\beta}$, $P_{3i}=0$ (where $i=(\a,3); \a=1,2$)  and 
\be
\te J_i(w) =k A^{(2)}_i=\frac{\ii }{4}k  \big( w^a\partial_i\bar w^a-\bar w^a\partial_iw^a\big)\ ,
 \la{45}
\ee
$A_i=A^{(2)}_i + A^{(4)}_i  + ...$   is the expansion of the  world-volume projection  of  the 
1-form $A$ in \rf{66}. 
The  fourth-order term  is 
\al{
& \te h^{(4)}_{ij}=\frac12|x|^4P_{ij}
 +\frac14|x|^2
 \big(\partial_ix\,\partial_j\bar x+\partial_jx\,\partial_i\bar x\big)
 -\frac18|w|^2
 \big(\partial_iw^a\partial_j\bar w^a
 +\partial_jw^a\partial_i\bar w^a\big)
 \no\\
 &\te -\frac18\big[(w^a\partial_i\bar w^a)
 (\bar w^b\partial_jw^b)+(i\leftrightarrow j)\big]
 +\frac1k\big(\delta_i^3A^{(4)}_j+\delta_j^3A^{(4)}_i\big)
 +A^{(2)}_iA^{(2)}_j\ ,
 \qquad  A^{(4)}_i=-\frac12|w|^2A^{(2)}_i\ .
 \la{47}
}
  The
quadratic  part of  the action  coming from the volume part in \rf{7}
 is \be \cL^{\rm (v)}_{2b}=\tfrac12 g^{ij}H_{ij}
=|\partial x|^2+2|x|^2+|\partial w|^2+J_3(w)\ , \ee
and the bosonic Wess--Zumino
term  in \rf{7} contributes 
\be
 \cL^{\rm (wz)}_{2b+4b}\te 
 =-\frac{3\ii k}{4}\,
 (x\partial_3\bar x-\bar x\partial_3x)
 \big(1+\frac32|x|^2\big)\ .
 \la{48}
\ee
The quadratic part of the  bosonic  Lagrangian is the same as in \rf{17}. 
The  quartic  bosonic vertices from the volume part of the action are   found to be 
\al{
 \cL_{4x}=&\te 2|x|^4+|x|^2|\partial x|^2
 -\frac12\big|g^{ij}\partial_ix\,\partial_jx\big|^2
\te  +\frac{k^2}{2}\,|x|^2\,\partial_3x\,\partial_3\bar x
 \la{49}\ , \\
 \cL_{2x,2w}={}&|\partial x|^2|\partial w|^2
 -|\partial x\cdot\partial w^a|^2
 -|\partial x\cdot\partial\bar w^a|^2
 \te +\frac{k^2}{2}\,|x|^2\,\partial_3w^a\partial_3\bar w^a
 \no\\
 &\te +J_3\big(|\partial x|^2+2|x|^2\big)
 -g^{\alpha\beta}
 \big(\partial_3x\,\partial_\alpha\bar x
 +\partial_3\bar x\,\partial_\alpha x\big)J_\beta
 -\frac{k^2}{2}\,|\partial_3x|^2J_3\ ,
 \la{50}\\
 \cL_{4w}={}&\te -\frac12|w|^2|\partial w|^2
 -\frac12(w^a\partial_i\bar w^a)(\bar w^b\partial^iw^b)
 +\frac12\big(|\partial w|^2\big)^2
 -\frac12(\partial_iw^a\partial^i\bar w^b)
 (\partial_j\bar w^a\partial^jw^b)
 \no\\
 &\te -\frac12\big|\partial_iw^a\partial^iw^b\big|^2
 -\frac12|w|^2J_3+|\partial w|^2J_3
 -g^{ij}J_j\big(\partial_3w^a\partial_i\bar w^a
 +\partial_iw^a\partial_3\bar w^a\big)\ .
 \la{51}
}
Adding the quartic part of \rf{48} gives 
\be
 \cL_{4b}=\cL_{4x}+\cL_{2x,2w}+\cL_{4w}
 -\tfrac{9\ii k}{8}\,|x|^2
 (x\partial_3\bar x-\bar x\partial_3x)\ .
 \la{52}
\ee
The  expectation value of \rf{52} is  computed in 
Appendix~\ref{apf}. 
Defining the
coincident limits of  the $\del_3$-differentiated Green's functions
($\G^{(3)}_\Phi\sim\langle\Phi\,\partial_3\bar\Phi\rangle$,
$\G^{(33)}_\Phi\sim\langle\partial_3\Phi\,\partial_3\bar\Phi\rangle$, $\Phi=(x, w^a)$, see
Appendix~\ref{apf})  one finds that $\langle\cL_{4b}\rangle$   takes a  square form 
\al{&\qquad \qquad \langle\cL_{4b}\rangle=-\tfrac38\,\mC^2\ \ , 
 \la{55}\ \qquad \qquad  \mC= \tfrac{k}{2}\Big[-2\G^{(3)}_x + 
 k\G^{(33)}_x
 +3\big(2\G^{(3)}_w+k\G^{(33)}_w\big)\Big]\ .
}

\subsection{Fermionic  sector}

The fermionic vertices are obtained from the all-order supervielbein and
super-three-form   expressions in  \ci{deWit:1998yu}, followed by the static and kappa
gauge fixing, see  Appendix~\ref{apa}.  The quadratic  fermionic part  was already
 given  in the 3d form in \rf{18}. 
  Here  we    present the  final form of $ \cL_{2b,2f}$  and $ \cL_{4f}$ and their  expectation values 
 with   details of the derivation  given in
Appendices~\ref{apc}--\ref{aph}.

Let $H_{ij}$ be the quadratic bosonic perturbation \rf{46} of the
induced metric, $\rho^{(r)}_i$ and $D^{(r)}_i$ the order-$r$ terms in the expansion of the 
pullback of Clifford algebra matrix and the Killing-spinor derivative, and
$\Gamma_{(r)}$ the corresponding terms in the $\kappa$-symmetry  projector (see
Appendix~\ref{apa}). 
 Then ($H\equiv  g^{ij} H_{ij}$)  
\al{
 \cL_{2b,2f}={}&
 \te  -\ha \big(H^{ij} - \frac12Hg^{ij}\big)
 \tb\rho_iD^{(0)}_j\theta
 +\frac12g^{ij}\tb\big[
 \rho^{(2)}_iD^{(0)}_j+\rho^{(1)}_iD^{(1)}_j
 +\rho_iD^{(2)}_j\big]\theta
 \no\\
 &\te -\frac14g^{ij}\tb\Gamma_{(1)}
 \big[\rho^{(1)}_iD^{(0)}_j+\rho_iD^{(1)}_j\big]\theta
 -\frac14g^{ij}\tb\Gamma_{(2)}\rho_iD^{(0)}_j\theta\ .
 \la{56}
}
  The explicit form of the 
$x^2\theta^2$, $xw\theta^2$ and $w^2\theta^2$  terms in 
\rf{56} is  given in Appendix~\ref{apc}. 

Taking the expectation  value of \rf{56}  (see Appendix~\ref{apg})
one finds that 
the $xw\theta^2$ term  vanishes.  In the $x^2\theta^2$ and $w^2\theta^2$
sectors all terms multiplying undifferentiated fermion bilinears reduce to
the bosonic equations of motion.  The remaining terms  combine  into  the cross
product of the bosonic combination $\mC $ in  \rf{55} and the  analogous fermionic
combination  $\mR$ 
\be
 \langle\cL_{2b,2f}\rangle
 =\tfrac34\,\mC\,\mR\ .
 \la{57}
\ee
 Here $\mR$  is defined as follows. 
 Let $\vT$   be a    label of each  fermionic  set   with masses  given in \rf{37}. 
 In terms of the coincident-limit  $\mathrm{AdS}_2$ fermion propagator coefficient $G_f$, one has 
 \al{
& \langle\theta_{\vT,n}\,\bar\theta_{\vT,-n}\rangle\Big|_{\s'\to\s}
 =-\Gb\,\Pi_\vT\,\G_f(m_{\vT,n})\ ,\ \qquad \qquad \Gb=\ii\Gamma_{1234}\ , \la{60a}\\ 
& \qquad \ \ \ 
 \mR=  \tfrac{k}{4\pi}\sum_{n=-\infty}^\infty
\sum_\vT   (m_{\vT,n}+1)\,\G_f(m_{\vT,n})\ .
 \la{60}
}
  $\Pi_\vT$ is a projector to  internal charges
 and the mass parameters $m_{\vT,n}$ form the
$3+3+1+1$ set in \rf{37}   (see Appendix \ref{apb}).  
$\theta$ is 11d  Majorana fermion  and  $\bar \theta_\vT$ 
  carries the   conjugate projector $\Pi_\vT^c$ (cf. 
\rf{39}).  
At $n=0$ the two massless singlet Majorana blocks ($m_{\T,n}\to \pm p_n\to0$) form a single complex Dirac fermion $\chi_0$. Accordingly, the two terms in the Majorana-block sum are expressed in terms of the same complex-Dirac propagator coefficient $G_f(0)$, so that their total contribution to \rf{60}  is $2G_f(0)$, 
defined in section~\ref{X-result}.



The  four-fermion part of the Lagrangian is  found to be (see Appendix \ref{apd})
\be
 \cL_{4f}=\te \frac1{192}\,g^{ij}\,\tb\rho_i(\cM^2)_+D^{(0)}_j\theta
 -\frac1{16}\,g^{il}g^{jk}B_{ij}B_{kl}
 +\frac1{16}\,\big(g^{ij}B_{ij}\big)^2\ , \qquad \qquad B_{ij}\equiv \tb\rho_iD^{(0)}_j\theta\ . 
 \la{59}
\ee
Here $(\cM^2)_+$ is the kappa-projected mass-squared matrix  given in 
Appendix~\ref{apa}.  The coefficient of the  first $\cM^2$ term receives equal contributions from both 
 the volume   and the Wess--Zumino  terms in the M2 brane action.\
Taking the expectation value of \rf{59} one finds that the result has a  similar  form as in  \rf{55}  and \rf{57}
(see Appendices~\ref{ape} and \ref{aph})
\be
 \langle\cL_{4f}\rangle=-\tfrac38\,\mR^2\ .
 \la{61}
\ee

 
\section{Vanishing  of the  2-loop correction} 
\label{X-result}

Combining \rf{55}, \rf{57} and
\rf{61} we find the  perfect square  expression 
\be
 \langle\cL_4\rangle
 =-\tfrac38\,\cX^2 \ , \qquad \qquad   
 \qquad
 \cX\equiv \mC-\mR\ ,
 \la{62}
\ee
where $\mC$ and $\mR$ are defined in \rf{55} and \rf{60}. 
Let us note that  $\cX$ in \rf{62}   can be obtained  as an expectation value  of a 
 particular quadratic operator  in the
free 3d theory in \rf{17},\rf{18}:\foot{The 
 operator $\mathcal{O}$ below is introduced simply to reproduce 
the expression for  $\cX$ in $\cX^2$ (that was obtained 
from  expectation values of the quartic bosonic, fermionic and mixed terms in Appendices \ref{apf},\ref{apg},\ref{aph})
 by directly   evaluating $\langle \cO\rangle$ in quadratic theory. A possible deeper meaning of $\cO$ 
  remains to be understood.}
\al{
 &\qquad \qquad  \qquad\qquad  \qquad \cX=\langle\cO\rangle \ , \la{64} \\
& \cO=2\bar x  P(P-1) x+2\bar w^a P(P+1) w^a +\bar\chi\,\ii\gamma^3\,(\hat P+1)\chi 
 + \bar\psi^a\,\ii\gamma^3\, \hat P\psi^a\ , \la{649}\\
 &\qquad  \qquad     P\equiv -\ii\partial_y= -\tfrac{\ii }{2}k \del_3 \ , \ \qquad\qquad   \hat P \equiv -\tfrac{\ii}{2}  \overleftrightarrow \del_y\  .
 \no
}
Expanding the fields or 
the coincident Green's functions in \rf{55} and
\rf{60} in Fourier modes gives  the infinite  sum
representation of $\cX$  in  \rf{22},\rf{23}.
To recall, 
only the two coincident propagator constants  
are required:
 the  bosonic 
one  $\G_b$ in \rf{20} and the fermionic one $\G_f$ in \rf{020} with the latter  defined as      \ci{Camporesi:1990wm,Camporesi:1995fb}
\be
 \big(\slashed\nabla+\ii\gamma^3m\big)^{-1}\Big|_{\s'\to\s}
 =-\ii\gamma^3\,\G_f(m)\ ,
 \qquad\qquad 
 \G_f(m)=-\G_f(-m)\ \ \ \   (m\neq0)\ .
 \la{67}
\ee
With the dimensions $\Delta_{x,n},\Delta_{w,n}$ in \rf{21} and fermion masses  in \rf{17b}  we get from \rf{55},\rf{60} or  \rf{64}
the expression in  \rf{22},\rf{23} ($p_n= \ha k n$; see also Appendix~\ref{api})
\al{
& \cX=\tfrac{k}{4\pi}\sum_{n=-\infty}^\infty\X_n\ ,
\la{680} \qquad\\
 \X_n=2p_n(p_n-1)\,\G_b(m_{x,n})&+6p_n(p_n+1)\,\G_b(m_{w,n})
 -2(p_n+1)\,\G_f(m_{\chi,n}) -6p_n\,\G_f(m_{\psi,n})\ .
 \la{68}
}
Combining  the  level $n$  and  $-n$ contributions for $n\ne 0$ 
 one finds  from  \rf{20},\rf{020} that  $\X_n + \X_{-n}$  is just an  $n$-independent constant $- {1\ov \pi}$  in \rf{24}.\foot{\vthree The 
factor of 2 in $2(p_n+1)\G_f(m_{\chi,n})$   should not be read  as the sum of the $\chi$ and $\chi^c$  blocks  each  
contributing  $\G_f$ with the same sign.  On conjugate blocks $\epsilon_{\T^c}=-\epsilon_\T$, so by \rf{78c} a single 
physical  boundary  condition  corresponds  to  opposite $\eta$,  and $\G_f$ is  correspondingly  odd, consistently with 
$m_{\T^c}=-m_\T$. 
 After the reindexing $n\to -n$, this is equivalent to the Majorana pairing $m_{\T^c,-n}=m_{\T,n}$ in \rf{39}.
The two contributions    agree  because the weight  generated by 
$\bar\chi\, \ii \g^3(\hat P +1)\chi$  in \rf{649}   reverses sign on the conjugate block as well; equivalently, \rf{68} 
is written after the reindexing  $n\to -n$  implied by \rf{39}.}

The universal pair value $\X_n+\X_{-n}=-{1\ov \pi}$ suggests assigning the corresponding half-pair value $\bar{\X}=-{1\ov 2\pi}$ to the unpaired zero level contribution $\X_0$
which   is determined just by the massless  fermion $\chi_0$    contribution
  \be 
  \X_0 =-2\,\G_f(0)= \bar \X \ , \qquad \qquad \te  \bar \X= - {1\ov 2\pi} \ . \la{70} \ee
  In this case we would  get as in \rf{26} 
  \be\la{700}
  \sum_{n=-\infty}^\infty \X_n =  \bar \X  \sum_{n=-\infty}^\infty 1 = \bar \X \big[1 + 2 \zeta (0) \big] = 0 \ . 
  \ee
    
  The definition of  $\G_f(0)$ is  potentially ambiguous (cf. \rf{25})    depending on how the limit $m \to 0$ is taken
  \ci{David:2023btq} (see also
\ci{Gripaios:2008rg}).
The   fermion field conformal dimension $\Delta_f=|m|+\ha$ is even in $m$, but 
 there 
 are two different  Dirac field   asymptotics  near the $\AdS_2$ boundary
$
 z^{\frac12+m}\ ,
 $ and $
 z^{\frac12-m} .
$
For $m\neq0$, the standard branch $\Delta_f=\frac12+|m|$ used in \rf{020}  selects the asymptotic behavior according to the sign of $m$. At $m=0$ the two exponents coincide, while the two possible boundary projectors remain distinct.

Equivalently, the alternative fermion quantization  corresponds to  $m\to-m$, under
which $\G_f$ is odd.  A 
definition of the  $m \to 0$   limit  corresponds  to a particular  choice of a  boundary condition.
This is the same type of boundary data  that  appears in the  discussions of 
 localization in  AdS$_p$  and  $\AdS_2\times S^1$
\ci{David:2016onq,David:2018pex,David:2019ocd}.

As in \ci{David:2023btq}  we may consider  the  massive  $\AdS_2$   fermion bilinear 
with the insertion of the  chirality matrix $\ii \gamma^3$. Using  point-splitting regularization one gets
(cf. eq.~(3.51) in \ci{David:2023btq})
\be
 \big\langle\bar\chi\, \ii \g^3\, \chi\big\rangle
 =\frac{m}{2\pi}
 \Big[-\log \Lambda^2 + \uppsi(|m|)+\uppsi(1+|m|)\Big]
 =\frac{m}{\pi}
 \Big[-\log \Lambda + \uppsi(|m|)+\frac1{2|m|}\Big]\ ,
 \la{77}
\ee
where we used that  $\uppsi(1+a)=\uppsi(a)+a^{-1}$ and  absorbed scheme-dependent  ($m$-independent) constants into  $\Lambda$. This is precisely  $-2\,\G_f(m)$
 in \rf{020}  and thus 
\be
 \big\langle\bar\chi\,\ii\gamma^3\chi\big\rangle
 =-2\,\G_f(m)\ , \qquad \ \ \ \big\langle\bar\chi\,\ii\gamma^3\chi\big\rangle_{m\to0^\pm}
 =\mp\frac1{2\pi}\ ,
 \qquad\qquad 
 \G_f(0^\pm)=\pm\frac1{4\pi}\ .
 \la{79}
\ee
The resulting  jump across $m=0$, i.e. 
$\big\langle\bar\chi\ii\gamma^3\chi\big\rangle_{0^+}-\big\langle\bar\chi\ii\gamma^3\chi\big\rangle_{0^-}=-\frac{1}{\pi}$,
was found in \ci{David:2023btq} both from the exact Green's function and 
from a direct numerical evaluation of the fermion determinant; 
it is a  feature of  $\AdS_2$  (being proportional to the inverse $\AdS_2$ radius)  
with no analog  in flat 2d space.\foot{A related observation
applies to the 1-loop   fermionic determinant.  
 The
renormalized fermionic log det  is, up to an $m$-independent constant, 
\ci{David:2023btq}
$
 {\cal F}_f(m)=-\int_0^\infty d\nu\,
 \big[\nu\coth(\pi\nu)\log(\nu^2+m^2)
 -\nu\log\nu^2\big]\ ,
 $
and since $\nu\coth(\pi\nu)\big|_{\nu\to 0} \to{1\ov \pi}$ 
one has $
 {\cal F}_f(m)={\cal F}_f(0)- |m|
 +\hbox{terms with continuous first derivative at $m=0$} 
$.
The determinant is thus even and continuous in $m$, but not differentiable:
the one-point function in  \rf{79} is a one-sided  $m$-derivative
of the $|m|$ cusp.  The  average of the two derivatives vanishes,
but it  does not correspond  to  either admissible  quantization, so the oddness
of $\G_f(m)$ away from the origin does not by itself imply  $\G_f(0)=0$.}

The invariant meaning of the two signs in \rf{79} is a choice of the  boundary  conditions for  
 the massless Dirac operator.  While at $m=0$ the two  asymptotics  $z^{{1\ov 2} \pm m}$   
degenerate,  they  remain  distinguished  by  the eigenvalue of 
\be 
 \uN \equiv \g^{\hat z}\,\Gba, \ \ \ \ \ \   \Gba \equiv   \ii\, \g^3 \ , \qquad \ \  
 \uN^2 = 1 \ , \qquad \ \  \{\uN,\g^{\hat z}\}= \{\uN,\Gba\}=\{\uN,\g^3\}=0 \ , 
 \la{78a}
\ee
where  $\g^{\hat z}$  is the gamma matrix along the unit normal to the $\AdS_2$  boundary  and 
$\Gba =\ii \g^3$  is the matrix  multiplying  the mass  in \rf{67}. Indeed, writing $\chi = z^{1/2}\tilde\chi$ 
and using  $\slashed\nabla  ...= z^{3/2}\slashed\partial (z^{-1/2}...)$  one gets, to leading radial order,
$z\del_z\tilde\chi = - m \uN \tilde\chi$, i.e. 
\be 
 \chi \sim z^{{1\ov 2} - m \nu} \ , \qquad \ \ \uN\chi = \nu \chi \ ,  \ \ \ \ \ \ \ \nu=\pm 1 \ , 
 \la{78b}
\ee
so that for $m\ne 0$ the branch $\Delta_\chi=\ha+|m|$ used in \rf{020} selects  $\nu = - {\rm sign}\, m$.  The two admissible 
massless  cases are then 
\be 
 \lim_{z\to0}z^{-1/2}\PPi_\eta\,\chi=0 \ , \qquad \qquad 
 \PPi_\eta = \ha \big( 1 + \eta\, \uN\big) \ , \qquad \ \ \eta = \pm 1 \ , 
 \la{78}
\ee
the surviving  boundary  component  carrying the value of   $\uN$ equal to $-\eta$.  Note that the grading is that of $\uN$  and 
not of $\g^3$:  since  $\{\uN,\g^3\}=0$, a  $\uN$  eigenspinor  is never a $\g^3$  eigenspinor.  
On a  charge block  $\Pi_\T$  of Appendix~\ref{apb}  one has, using  $\Gba \to \Gb = \ii \Gamma_{1234}$ and 
$\ss_0\big|_{\Pi_\T}=\ss_1\ss_2\ss_3 \big|_{\Pi_\T} = \epsilon_\T$, 
\be 
 \Gamma_1\Gamma_2\Gb = -\ii \Gamma_{34} = -\ss_0 \ , 
 \qquad \qquad 
 \uN\big|_{\Pi_\T} = -\, \g^{\hat t} \ , 
 \la{78c}
\ee
so that the projector  is built  from the  tangential  gamma matrix.  
The labels $\pm$  here depend  on  the  orientation  and  $\gamma$-matrix conventions.  
The two  cases  in \rf{78}  are mapped  into each other  by  $\gamma^3\to-\gamma^3$ 
(equivalently, by  $m \to -m$),  under which  $\uN$  is odd. Thus   no property of the free  massless   fermion  $\chi_0$   by itself  
can distinguish  between them:  the choice  must be   correlated 
with the boundary conditions of the other fields, i.e. fixed  by  a symmetry  that relates them. 

For supersymmetric theories on $\AdS_2\times S^1$  the required  criterion  
was  found in \ci{David:2018pex}  in the example of  the  chiral multiplet 
(see \ci{David:2019ocd}  for a  general analysis). 
There  one compares two  assignments of  the $\AdS_2$   boundary conditions for  the  KK  modes: 
(i) the normalizable  ones, imposed  mode by mode  separately on the  bosons and the fermions,  
and (ii) the supersymmetric ones,   obtained by demanding that  the space  of 
allowed field configurations  closes under the  preserved supercharge $Q$.  
 Since $Q\psi \sim \gamma^i \partial_i\phi\, \varepsilon+...$, 
the  boundary  falloff  of the fermion  is  inherited from  that of its scalar partner, so that 
supersymmetry  {\it correlates}  the fermion quantization  with the branch chosen for the scalar. 

The two assignments  coincide  for  generic  masses  but  differ  whenever a KK level falls into 
a   range   bounded by  the points where  the scalar  mass  saturates the  BF bound and where the 
fermion mass  vanishes.  
   In the latter case  it is the supersymmetric assignment  that is 
consistent  \ci{David:2018pex,David:2019ocd}.


In the present case, the supersymmetric boundary conditions follow directly from the discussion in 
section~\ref{quadratic}.
According to \eqref{sk4},
at the $n=0$ level the massless fermion $\chi^c_0$ belongs to the same
multiplet as the massless scalar $\bar w^3_0$. For the scalars the boundary
conditions are not ambiguous: since the minimal surface is localized at a
point of $\CP^3$, the fluctuations $w^a_0$ must obey the Dirichlet boundary
condition corresponding to $\Delta_w=1$ ($w\sim\beta\,z$ as $z\to0$).

For the pair $(\bar w^3_0,\chi^c_0)$ in \rf{sk4}, the Dirichlet
condition on $\bar w^3_0$, together with the conjugate condition in
\rf{sk5}, gives   
\be
 \lim_{z\to0}z^{-1/2}\PPi_+\chi_0=0\ .
\ee
Hence the supersymmetric case is $\nu=-1$, i.e. $\eta=+1$ in \rf{78},
and by the coincident limit of the Green's function computed in
Appendix~\ref{api} (see \rf{i34})
\be
 \G_f(0)\big|_{\rm susy}={\eta\,\epsilon_\T\ov4\pi}\ ,
 \la{773}
\ee
where $\epsilon_\chi=+1$ for the mode $\chi_0$. 
The massless point must therefore be approached from the $m\to0^+$ side,
i.e. the supersymmetric quantization of $\chi_0$ is the one corresponding
to the $m\to0^+$ limit in \rf{020},\rf{79}. Thus
\be
 \G_f(0)\Big|_{\chi_0}=\G_f(0^+)=\tfrac{1}{4\pi}\ .
 \la{781}
\ee
This is the choice used in \rf{25},\rf{70} that gives $\X_0=\bar\X$
and thus $\cX=0$ in \rf{700}, for $k>2$.

Let us note  that the  two cases  in \rf{78}  can  also be  distinguished  by   separated-point   correlators  of the 
boundary  operators:  both  have  $\Delta_f=\ha$  and thus the same  conformal power law, but the external fermions carry 
either $\PPi_+$ or $\PPi_-$ projectors,  so   chirality-odd  structures  (like the bilinear one in \rf{649}) 
 change  sign  between the two.  Such 
correlators   diagnose  which  boundary condition  has been imposed  but do not  by themselves  select it. \vthree The  argument 
leading to \rf{773}  is the linearized  form of the corresponding  supersymmetry  Ward identity;  its  extension  to  
separated-point   fermionic  defect  correlators  and  those of  $w^3_0$  with its  fixed  Dirichlet  boundary condition  
would  provide a  further  test.

\section{Conclusions}
\label{discussion}

As   already discussed in the Introduction, the  result  that $\cX$ in \rf{680},\rf{700}  and thus the 2-loop correction 
$f_2$ in \rf{22}  vanishes\foot{Above 
we assumed that $k>2$ but as  discussed in Appendix \ref{apj}  $\cX$  vanishes also for $k=1,2$.}
  matches the  localization result (for $k>2$)  in the grand-canonical ensemble  form. 
 This  provides a highly non-trivial test  of  the proposal of \ci{Gautason:2025per,Gautason:2025plx}.  

This  vanishing result   follows from:

(i)  organization of the 2-loop correction into a perfect square in \rf{62};

 (ii) cancellation of all transcendental $k$-dependent parts   between the bosonic and fermionic parts  in
 
 \qquad  the summand of  
 $\cX$ in \rf{23},\rf{24};
 
  (iii)   \vfour  definition of the   massless level $n=0$   fermion 
 contribution  in \rf{24}   using  the  supersymmetric 
 
  \qquad boundary  domain  \rf{78},\rf{773}, 
  equivalently  the $m \to 0^+$ limit   in  \rf{25},\rf{79}.   
 
 \noindent  
 Like (iii),  the first two   facts    should also  be a consequence  of the 3d   supersymmetry of the  quadratic fluctuation   action in \rf{17},\rf{18}   extended to the quartic interaction  part of  the M2 brane action. 
 Its origin is a non-linearly realised   supersymmetry  present in the M2  brane action after fixing the static and 
 $\kappa$-symmetry gauge (see, e.g.,   \ci{Tseytlin:2025dae}  for a review and references). 
 Here the perfect square structure in \rf{62} was obtained by explicit evaluation of the quartic vertices. Making this  underlying supersymmetry  manifest on the $\AdS_2 \times S^1$  background  beyond the quadratic fluctuation  order  is an interesting open problem that could elucidate the perfect square structure  obtained above.
 
Another   2-loop computation for   M2 brane in $\AdS_4 \times S^7/\ZZ_k$ 
analogous to the present one is for the  $S^3/\ZZ_k$ instanton 
 with 
  the 1-loop contribution found  in  \ci{Beccaria:2023ujc} (see also \ci{Kurlyand:2026yke}). 
 The corresponding M2 brane partition function  represents a leading in large $N$ 
  non-perturbative contribution to the ABJM free energy,
    and  the 2-loop correction  is   also expected to   vanish   and  to  match the localization  prediction  in the grand canonical  ensemble.
    This 2-loop instanton   case was  recently discussed   in \ci{Tseytlin:2026dhy}.

\section*{Acknowledgements}
We   thank L. Wulff  for  useful comments. 
MB is supported by the INFN grant GAST.  
SK is  supported  by   the President's PhD Scholarship at Imperial College London.
 AAT and JvM are  supported by the STFC Consolidated Grant    ST/X000575/1 
 and  STFC Consolidated Award UKRI4230: Abdus Salam Centre for Theoretical Physics at Imperial College London.
 Some of the computations of this paper were done using 
 ChatGPT models   and checked  using Claude  models.




\appendix


\section{M2  brane action}\label{apa}

The starting point is the general covariant supermembrane action in an
11d supergravity background, in Euclidean signature
\ci{Bergshoeff:1987cm}
\begin{equation}
 S_E=T_2\int_{\Sigma_3}\dd^3\sigma\,\Big[
 \sqrt{ \hat h(Z)}
 -\tfrac{\ii}{3!}\,\epsilon^{ijk}\,
 E_i^{\widehat A}E_j^{\widehat B}E_k^{\widehat C}\,
 \hat C_{\widehat C\widehat B\widehat A}(Z)\Big]\ ,
 \qquad
 \hat h_{ij}=E_i^{A}E_j^{B}\delta_{AB}\ ,
 \la{114}
\end{equation}
where $Z^\MM=(X^M,\Theta^\mu)$ are local coordinates on superspace,
$
 E_i^{\widehat A}=\partial_iZ^\MM\,E_\MM{}^{\widehat A}(Z),
 $   $
 \widehat A=(A,\alpha)$, 
are the pullbacks of the supervielbein one-forms. The induced metric
$\hat h_{ij}$ is built from their vector components only, and
$\hat C_{\widehat A\widehat B\widehat C}$ are the superspace components of the
three-form whose bosonic part is $C_3$ in \rf{6},\,\rf{7}.  For a background
satisfying the 11d supergravity equations the action is
invariant under the kappa symmetry
\begin{equation}
\delta_\kappa Z^\MM E_\MM{}^{A}=0, \qquad \delta_\kappa Z^\MM E_\MM{}^{\alpha}=\big[(1+\Gamma)\kappa(\sigma)\big]^\alpha  , \ \ \qquad
 \Gamma=\tfrac{\ii}{3!\sqrt{\hat h}}\,\epsilon^{ijk}
 E_i^{A}E_j^{B}E_k^{C}\,\Gamma_{ABC} ,
 \qquad
 \Gamma^2=1\ .
 \la{378}
\end{equation}
At $\theta=0$ the action \rf{114} reduces to \rf{7}.
The matrix \rf{378} evaluated on the bosonic embedding is  $\Gamma_{\rm B}$ \rf{225}, and its
background value is the constant projector $\Gk=\ii\Gamma_{12,11}$ of
\rf{119}.

The  $\AdS_4 \times S^7/\ZZ_k$  target space metric is
\begin{align}
 \dd s^2&= R^2\big(\tfrac{1}{4}\dd s^2_{\AdS_4}
          +\dd s^2_{S^7/\ZZ_k}\big)\ ,\qquad 
 \dd s^2_{\AdS_4}
 =\frac{(1+{1\ov 4} X^2)^2}{(1-{1\ov 4} X^2)^2}\dd s^2_{\AdS_2}
 +\frac{\dd X^p\dd X^p}{(1-{1\ov 4} X^2)^2},\qquad p=1,2,
 \la{106}\\
 \dd s^2_{S^7/\ZZ_k}
 &=\dd s^2_{\CP^3}+\tfrac1{k^2}(\dd \varphi+kA)^2,
 \la{107}\\
 \dd s^2_{\CP^3}
 &=\frac{(1+|W|^2)\dd W^a\dd\bar W^a
 -W^a\bar W^b\dd\bar W^a\dd W^b}{(1+|W|^2)^2},
 \qquad
 A=\tfrac{\ii}{2}\frac{W^a\dd\bar W^a-\bar W^a\dd W^a}{1+|W|^2}.
 \la{108}
\end{align}
The embedding of the M2  brane  into  $\AdS_4 \times S^7/\ZZ_k$   identifies $\s^1,\s^2$ with $\AdS_2$ coordinates inside 
$\AdS_4$ and $\varphi$ with $\sigma^3$, with 
$X^p=W^a=0$.
The induced 3d metric is 
\be
 h^{(0)}_{ij}=\frac14g_{ij},\qquad
 g_{ij}=\begin{pmatrix}g^{\AdS_2}_{\alpha\beta}&0\\[1mm]0&4/k^2\end{pmatrix}\ ,
 \la{109} \qquad 
 \sqrt{h^{(0)}}=\frac18\sqrt g
 =\frac1{4k}\sqrt{g_{_{\AdS_2}}} \ . 
 \ee
The geometric coordinates and the superspace spinor are related to the
canonically normalised  3d fluctuation  fields as
\begin{equation}
 X^p=\sqrt2\,x^p,\qquad W^a=\frac1{\sqrt2}w^a,
 \qquad \Theta=\frac1{2\sqrt2}\theta\ , \qquad x\equiv x^1+\ii x^2\ . 
 \la{111}
\end{equation}
The tangent $\AdS_2$ directions are labelled by $\alpha,\beta=1,2$.
The two transverse coordinates are $x^p$, $p=1,2$.    The $\CP^3$ tangent directions
are $I,J=5,\ldots,10$, and the Hopf direction is $11$.

\iffa 
The coefficient naturally accompanying the fluctuation Lagrangian written
with the auxiliary metric $g$ is
\begin{equation}
 \tau=\frac{\rT_2}{8}.
 \la{113}
\end{equation}
Before gauge fixing it is useful to keep the Euclidean three-form coefficient
explicit,
\begin{equation}
 S_E=8\tau\int_{\Sigma_3}\sqrt{\det h}
 +\varkappa\int_{\Sigma_3}B_3 .
 \la{114}
\end{equation}
The kappa-projector argument in \S\ref{signchain} fixes
$\varkappa=8\ii\tau$; this sign is not an independent convention once
\rf{115} and \rf{119} are chosen.  Before the final loop
rescaling, the action is
$S_E=S_{\rm cl}+\tau\int\sqrt g(\cL_2+\cL_4+\cdots)$.
After $x,w,\theta\mapsto\tau^{-1/2}(x,w,\theta)$, every quartic vertex
carries a factor $\tau^{-1}$.
\fi 

We use hermitian Euclidean gamma matrices
\begin{equation}
 \{\Gamma_A,\Gamma_B\}=2\delta_{AB},\qquad
 \Gamma_{1\cdots 11}=\ii,\qquad \Gamma_*=\Gamma_{1234},\qquad \Gb=\ii\Gamma_* \ , 
\end{equation}
with $A,B$     having  components $\hat a,\hat b=1,\ldots,4$ along $\AdS_4$ and
$a',b'=5,\ldots,11$ along $S^7$.
The orientation is fixed   so that
\begin{equation}
 F_4=dC_3=\tfrac{3\ii}{8}R^3 \vol(\AdS_4)\ , \qquad \qquad F_{\hat a\hat b\hat c\hat d}=6\ii
 \epsilon_{\hat a\hat b\hat c\hat d} \ , 
 \la{115}
\end{equation}
where tangent-space coordinates  along $\AdS_4$ include  inverse  vielbeins    with factors of $2/R$ (cf. \rf{106}). 
Note that the expansion of the   corresponding $C_3$ potential   is  given by 
\begin{equation}
 C_3=\te \tfrac{3\ii}{16}\,\vol(\AdS_2)\wedge
 (X^1\dd X^2-X^2\dd X^1)
 \big[1+\frac34X^2+\OO(X^4)\big] . \la{138}
\end{equation}
Using \rf{109},\rf{111}  the WZ  term
$\int C_3$   leads to \rf{48}.

Then  the Killing-spinor derivative is
\begin{equation}
 \te D=\dd+\frac14\Omega^{AB}\Gamma_{AB}
 -\ii\big(E^{\hat a}\Gamma_{\hat a}
 -\frac12E^{a'}\Gamma_{a'}\big)\Gamma_* .
 \la{116}
\end{equation}
The supervielbeins of Ref.~\cite{deWit:1998yu} have the
expansion\foot{Direct contraction of the general 11d matrix 
$T_A{}^{BCDE}=\frac{1}{288}\big(\Gamma_A{}^{BCDE}
 -8\,\delta_A^{[B}\Gamma^{CDE]}\big)$
 with the 4-form flux
gives $
 T_{\hat a}{}^{BCDE}F_{BCDE}=-\ii\Gamma_{\hat a}\Gamma_*,
 $  and $
 T_{a'}{}^{BCDE}F_{BCDE}=\frac{\ii}{2}\Gamma_{a'}\Gamma_*
$.
}
\al{
  &E^A(X,\Theta)
 =e^A+2\sum_{m=0}^{15}\tfrac1{(2m+2)!}
 \bar\Theta\Gamma^A[\cM^2(\Theta)]^m D\Theta  \ , \\
 &  E^{\alpha}(X,\Theta)
 =\sum_{m=0}^{16}\tfrac{1}{(2m+1)!}[\cM^2(\Theta)]^m\, D\Theta^\alpha \ . 
\la{1166}
}
Here the  spinor bilinear $\cM^2$ is a  function of $\Theta$;  after the rescaling in \rf{111} one has 
\be
\cM^2(\Theta)=\tfrac18\cM^2(\theta) \ , \qquad 
 \cM^2(\theta) =\Gb\big(
 2\Gamma_{\hat a}\theta\tb\Gamma^{\hat a}
 +\Gamma_{a'}\theta\tb\Gamma^{a'}\big)
 +\big(
 \Gamma_{\hat a\hat b}\theta\tb\Gamma^{\hat a\hat b}
 -\tfrac12\Gamma_{a'b'}\theta\tb\Gamma^{a'b'}
 \big)\Gb \ . 
 \la{118}
\ee
We used that   $24\Gamma_{\hat c\hat d}F^{\hat a\hat b\hat c\hat d}
 =-288\ii\Gamma^{\hat a\hat b}\Gamma_* \ , 
 \Gamma^{a'b'\hat a\hat b\hat c\hat d}
 F_{\hat a\hat b\hat c\hat d}
 =144\ii\Gamma^{a'b'}\Gamma_* 
$.
Below $\cM^2$  will stand for   $\cM^2(\theta)$. Then 
\begin{equation}\te 
 E^A=e^A+\frac18\tb\Gamma^AD\theta
 +\frac1{768}\tb\Gamma^A\cM^2D\theta+\OO(\theta^6).
 \la{117}
\end{equation}
  The  $\kappa$-symmetry gauge is defined as  
\begin{equation}
P_-\theta=\theta,\qquad \tb P_-=\tb \ , \qquad \qquad P_-=\tfrac12(1-\Gk)\ , \qquad \qquad 
 \Gk\equiv \ii\Gamma_{12,11}\ . 
 \la{119}
\end{equation}
The matrices $\Gamma_\alpha$, $\Gamma_{11}$ and $\Gb$ commute with $\Gk$,
whereas $\Gamma_p$ and $\Gamma_I$ anticommute with it.\foot{We  use  $\Gamma_p$   as  a
shorthand for the associated $\AdS_4$ tangent space  gamma matrix $\Gamma_{p+2}$; thus the
actual transverse tangent labels of it  are $3,4$.}
  Hence the block of $\cM^2$ in 
\rf{118} contributing between gauge-fixed spinors is
\begin{align}
 (\cM^2)_+={}&\Gb\big(
 2\Gamma_\alpha\theta\tb\Gamma^\alpha
 +\Gamma_{11}\theta\tb\Gamma^{11}\big)
 +\big(
 \Gamma_{\alpha\beta}\theta\tb\Gamma^{\alpha\beta}
 +\Gamma_{pq}\theta\tb\Gamma^{pq}
 -\tfrac12\Gamma_{IJ}\theta\tb\Gamma^{IJ}
 \big)\Gb .
 \la{121}
\end{align}

\section{Quadratic fermionic  action}\label{apb}

Let us define the world-volume  Clifford algebra   matrices for the auxiliary metric  $g_{ij}$  as 
\begin{equation}
 \rho_\alpha=\te e_\alpha{}^a\Gamma_a,\qquad
 \rho_3=\frac2k\Gamma_{11},\qquad
 \{\rho_i,\rho_j\}=2g_{ij}.
 \la{122}
\end{equation}
The  covariant  derivative components  following from \rf{116} are\foot{Note that  for $\CP^3$ directions  
$dE^a + \Omega^a{}_{b} \wedge  E^b=0, \ \ \ 
 \Omega^a{}_b\big|_{\CP^3}=\frac12(W^a\dd\bar W_b-\bar W_b\dd W^a)
 +\frac12\delta^a_b(W^c\dd\bar W_c-\bar W_c\dd W^c).
$} 
\al{
 &\te D_\alpha^{(0)}=\nabla_\alpha-\frac12\rho_\alpha\Gb,
 \la{124}\qquad \qquad 
 D_3^{(0)}=\partial_3+\frac{\ii}{k}\widehat\xi
 +\frac1{2k}\Gamma_{11}\Gb ,\\
 &\te \widehat\xi=\frac12\ii
 (\Gamma_{56}+\Gamma_{78}+\Gamma_{9\,10}),\qquad 
 \qquad \xi=\big\{\pm\frac32,\pm\frac12\big\}.
 \la{123}
}
Here $\widehat\xi$ is an  effective charge operator and $\xi$  stands for its eigenvalue.
The quadratic part of the fermionic action is\foot{For a general rescaling $\Theta=a\theta$, the Nambu and Wess--Zumino
quadratic terms contribute $2a^2\tb\mathscr D\theta$ each.  Their sum is
$4a^2\tb\mathscr D\theta$, so the Majorana normalisation
$\frac12\tb\mathscr D\theta$ fixes $a^2=1/8$, as used in
\rf{111}.}  
\begin{equation}\te
 \cL_{2f}=\frac12\tb\mathscr D\theta,\qquad
 \mathscr D=g^{ij}\rho_iD_j^{(0)} \ , \qquad \qquad \mathscr D=\Dslash+\frac{k}{2}\Gamma_{11}\partial_3
 -\frac34\Gb+\frac{\ii}{2}\widehat\xi\, \Gamma_{11},
 \la{126}
\end{equation}
where $\Dslash$ is the $\AdS_2$ Dirac operator. 
On the $n$th Fourier mode in $\s^3$  we have 
\begin{equation}\te
 \mathscr D_n=\Dslash-\frac34\Gb
 +\frac{\ii}{2}(kn+\widehat\xi)\Gamma_{11}.
 \la{128}
\end{equation}
Let us  introduce 3 commuting charge projectors in 3 planes  and their product 
\al{
 &P_s^{(r)}=\tfrac12\big(1+s\,\ii
 \Gamma_{2r+3,\,2r+4}\big),\qquad s=\pm1,\qquad \ \  r=1,2,3 \ ,  
 \la{129}
\\
& \Pi_\T=P_-P_{s_1}^{(1)}P_{s_2}^{(2)}P_{s_3}^{(3)},
 \qquad \sum_\T\Pi_\T=P_-,   \qquad \qquad \T\equiv (s_1,s_2,s_3) \ , 
 \la{130}
}
where $P_-$ was defined in \rf{119}. 
Each $\Pi_\T$ has rank two, corresponding to selection of one 2d spinor.
Charge conjugation reverses the  charges in 3 planes  while leaving the $\kappa$-gauge
projector $P_-$ invariant  ($C$ is the 32$\times$32  Majorana  charge conjugation matrix; superscript $T$ denotes  the  transposition)
\begin{equation}
 C^{-1}\big(P_s^{(r)}\big)^T C=P_{-s}^{(r)},
 \qquad C^{-1}P_-^TC=P_-,
 \qquad C^{-1}\Pi_\T^TC=\Pi_{\T^c},
 \qquad \T^c\equiv(-s_1,-s_2,-s_3).
 \la{131}
\end{equation}
Then 
\begin{equation}
\widehat\xi\,\Pi_\T=\xi_\T\Pi_\T,
 \qquad\qquad 
 \Gamma_{11}\Pi_\T=\epsilon_\T\Gamma_*\Pi_\T, \qquad \ \ \ \  \xi_\T=\tfrac12(s_1+s_2+s_3),\qquad\qquad 
 \epsilon_\T\equiv s_1s_2s_3,
 \la{132}
\end{equation}
We used that 
$\Gamma_{1\cdots11}=\ii$ and $\Gamma_\kappa\Pi_\T=-\Pi_\T$.
Equivalently, if
\begin{equation}\la{b10}
 \ss_0=\ii\Gamma_{34},\qquad
 \ss_1=\ii\Gamma_{56},\qquad
 \ss_2=\ii\Gamma_{78},\qquad
 \ss_3=\ii\Gamma_{9\,10},
\end{equation}
then $\ss_0\Pi_\T =\ss_1\ss_2\ss_3\Pi_\T =\epsilon_\T \Pi_\T$ under  the $\kappa$-gauge  projector  and
$\Gb\theta=\ii\epsilon_\T\Gamma_{11}\theta$.

Let us define   the 2d mass matrix  $ \mathsf M_n$ so that 
\al{
&  \mathscr D_n=\Dslash+\tfrac12\mathsf M_n\ , \qquad 
  \mathsf M_n=-\tfrac32\Gb+\ii(kn+\widehat\xi)\Gamma_{11} \ , 
 \qquad  \mathscr D_{\T,n}=\Dslash+m_{\T,n}\Gb,
 \la{134}
\\
&\Gb^2=-1,\qquad \{\Gb,\Gamma_a\}=0,  \qquad  \mathsf M_n\Pi_\T=2m_{\T,n}\Gb\Pi_\T,
 \qquad
 m_{\T,n}=\tfrac12\big[\epsilon_\T(kn+\xi_\T)-\tfrac32\big]\ . 
 \la{135}
}
The fermion-tower mass parameters are then  (cf. \rf{37})
\begin{center}
\begin{tabular}{ccc}
\toprule
$(\xi_\T,\epsilon_\T)$ & multiplicity & $m_{\T,n}$\\
\midrule
$(-\tfrac12,+1)$ & $3$ & $\tfrac{kn}{2}-1$\\
$(+\tfrac12,-1)$ & $3$ & $-\tfrac{kn}{2}-1$\\
$(+\tfrac32,+1)$ & $1$ & $+\tfrac{kn}{2}$\\
$(-\tfrac32,-1)$ & $1$ & $-\tfrac{kn}{2}$\\
\bottomrule
\end{tabular}
\end{center}
This is the form that  is to  be used in a fermion propagator.
The shorter $3+3+2$ description
with masses $\ha kn\pm1$ and $\ha kn$ refers only to the squared-mass counting,
where the signs of the second triplet and one singlet are immaterial.
Note that  the bosonic and fermionic masses in \rf{17b} satisfy 
\al{
& \no
\sum_b m^2=2m_{x,n}^2+6m_{w,n}^2=2k^2n^2+4\ ,\qquad \quad 
 \sum_f m^2={\te 3\big(\frac{kn}{2}-1\big)^2
 +3\big(\frac{kn}{2}+1\big)^2
 +2\big(\frac{kn}{2}\big)^2}=2k^2n^2+6\ ,\\
 &\qquad \qquad \qquad  \qquad \qquad  \qquad  \sum_b m^2-\sum_f m^2=-2 \ . \la{b14}
}


\section{Boson-fermion part of   quartic  fluctuation action}\label{apc}

After the spinor rescaling in \rf{111}, the exact part of
the gauge-fixed  M2 brane action \rf{114}  which is quadratic in $\theta$ is given by 
\begin{equation}
 \cL_{\theta^2}=
 \frac{\sqrt{h}}{\sqrt g}\  h^{ij}
 \tb(1-\Gamma_{\rm B})E_i^A\Gamma_AD_j\theta \ , \qquad \qquad 
 \Gamma_{\rm B}=\frac{\ii}{3!\sqrt{h}}\,
\epsilon^{ijk}E_i^AE_j^BE_k^C\,\Gamma_{ABC}
=\Gk+\Gamma_{(1)}+\Gamma_{(2)}+...\ . 
 \la{225}
\end{equation}
where $h_{ij}$   stands for the bosonic part of the  projected metric  and $\Gamma_{(r)}$ are terms in expansion near classical solution. 
Expanding \rf{225} to second order in the bosonic fluctuations gives the  mixed  quartic vertex  quoted  
  in \rf{56}. 

In the $\kappa$-gauge \rf{119} the action has no cubic interaction   vertex.  Indeed,  given a   spinor bilinear $Y$ 
we may  define  $\Gk Y\Gk=\eta_Y Y$. Then the relevant parities are
$
 \eta_{\rho^{(0)}}=\eta_{D^{(0)}}=
 \eta_{\rho^{(2)}}=\eta_{D^{(2)}}=
 \eta_{\Gamma_{(2)}}=+1,
 \ \ 
 \eta_{\rho^{(1)}}=\eta_{D^{(1)}}=
 \eta_{\Gamma_{(1)}}=-1, 
$  so that the only non-zero  fermion-boson terms  start at quadratic order in both  fermions  and  bosons (see   below).


To distinguish the three complex $\CP^3$ directions from the $\AdS_2$
tangent index $a$, define, for $a=1,2,3$,
\begin{equation}\te
 \Gamma_a^{\rm CP}
 =\frac12\big(\Gamma_{2a+3}-\ii\Gamma_{2a+4}\big),
 \quad
 \bar\Gamma_a^{\rm CP}
 =\frac12\big(\Gamma_{2a+3}+\ii\Gamma_{2a+4}\big), \quad 
 \la{184}
 \{\Gamma_a^{\rm CP},\bar\Gamma_b^{\rm CP}\}=\delta_{ab},
 \quad
 \{\Gamma_a^{\rm CP},\Gamma_b^{\rm CP}\}
 =\{\bar\Gamma_a^{\rm CP}, \bar\Gamma_b^{\rm CP}\}=0.
\end{equation}
Each $\Gamma_a^{\rm CP}$ or $\bar\Gamma_a^{\rm CP}$ anticommutes with the
corresponding charge operator $\ii\Gamma_{2a+3,2a+4}$ and commutes with the
other two.  It therefore flips the relevant entry $s_a$ of the tower label.
  The notation $\Gamma_a$ without CP-superscript  below is
reserved for an $\AdS_2$ tangent gamma matrix.

Expanding in powers of bosons we have\foot{To see why no separate $\ii A^{(2)}\widehat\xi$ occurs, note that the unitary-frame $\CP^3$ connection is
given by $  \Omega^a{}_{b,(2)}\big|_{\CP^3} =\frac14(w^a\dd\bar w_b-\bar w_b\dd w^a) -\ii\delta^a_bA^{(2)}$. 
Adding the Hopf contribution
$\ii\delta^a_b({1\ov k} \dd \s_3+A^{(2)})$ cancels the last term.
  Its background ${1\ov k} \dd \s_3$ part produces
${\ii\ov k}\widehat\xi $ in $D_3^{(0)}$, while the fluctuating $A^{(2)}$ part is
already cancelled.  Including it again in \rf{189} would double count it.} 
\al{
 &2E_i^A\Gamma_A=\rho_i+\rho_i^{(1)}+\rho_i^{(2)}+...\ , \qquad \qquad  D=D^{(0)}+D^{(1)}+D^{(2)}+ ... \ , 
 \la{185}\\
& \rho_i^{(1)}=\sqrt2\big(
 \partial_ix^p\Gamma_p+\partial_iw^a\Gamma_a^{\rm CP}
 +\partial_i\bar w^a\bar\Gamma_a^{\rm CP}\big),
\qquad 
 \rho_i^{(2)}=|x|^2P_i{}^j\rho_j+2A_i^{(2)}\Gamma_{11}, \\
&
 \te D^{(1)}=\frac1{\sqrt2}e^\alpha x^p\Gamma_{\alpha p}
 -\frac1{\sqrt2}\dd x^p\Gamma_p\Gb
 -\frac{\ii}{2\sqrt2}\Gamma_{11}
 (\Gamma_a^{\rm CP}\dd w^a-\bar\Gamma_a^{\rm CP}\dd\bar w^a)
 +\frac1{2\sqrt2}
 (\Gamma_a^{\rm CP}\dd w^a+\bar\Gamma_a^{\rm CP}\dd\bar w^a)\Gb, \\
&\te
 D^{(2)}=-\frac12x^p\dd x^q\Gamma_{pq}
 -\frac12|x|^2e^\alpha\Gamma_\alpha\Gb
 +Q^{(2)}+\frac12A^{(2)}\Gamma_{11}\Gb , 
 \la{189}\\
&A^{(2)}=A_i^{(2)}\dd\sigma^i, \qquad 
 Q^{(2)}=\tfrac14\Omega_{(2)}^{IJ}\Gamma_{IJ},
 \qquad
 \Omega^a{}_{b,(2)}\Big|_{S^7}
 =\tfrac14(w^a\dd\bar w_b-\bar w_b\dd w^a).\la{1889}
}
In $\ \Gamma_{\rm B}$ in \rf{225} we have ($\rho_i^{(r)}=\rho_i^{(r)A}\Gamma_A, \ \ H= g^{ij} H_{ij}$)
\begin{align}\te
 \Gamma_{(1)}=\frac{\ii}{2\sqrt g}\epsilon^{ijk}
 \rho_i^{(1)A}\rho_j^B\rho_k^C\Gamma_{ABC},
\quad
 \Gamma_{(2)}=-\frac12H\Gk
 +\frac{\ii}{2\sqrt g}\epsilon^{ijk}
 \big(\rho_i^{(2)A}\rho_j^B\rho_k^C
 +\rho_i^{(1)A}\rho_j^{(1)B}\rho_k^C\big)\Gamma_{ABC}.
 \la{193}
\end{align}
Here $
 \{\Gamma_{(1)},\Gk\}=0,\   [\Gamma_{(2)},\Gk]=0,
$
while $D^{(1)}$ and $\rho^{(1)}$ are $\Gk$  odd and
$D^{(0)},D^{(2)},\rho^{(0)},\rho^{(2)}$ are $\Gk$ even.

Defining separate $x$ and $w$ parts of the quadratic metric perturbation in \rf{46} we have 
\begin{align}
 H^x_{ij}=\partial_ix\partial_j\bar x
 +\partial_jx\partial_i\bar x+2|x|^2P_{ij},
\qquad \qquad 
 H^w_{ij}=\partial_iw^a\partial_j\bar w^a
 +\partial_jw^a\partial_i\bar w^a
 +\tfrac4{k^2}(\delta_i^3J_j+\delta_j^3J_i).
 \la{228}
\end{align}
Also, we get  
\begin{align}
&\rho_i^{(1)}=R_i^x+R_i^w,
 \quad \rho_i^{(2)}=R_i^{x^2}+R_i^{w^2},
 \quad D_i^{(1)}=U_i^x+U_i^w,
 \quad D_i^{(2)}=U_i^{x^2}+U_i^{w^2},
 \la{234}
\\
 &R_i^x=\sqrt2\,\partial_ix^p\Gamma_p,
 \qquad R_i^w=\sqrt2(\partial_iw^a\Gamma_a^{\rm CP}
 +\partial_i\bar w^a\bar\Gamma_a^{\rm CP}),\qquad 
 R_i^{x^2}=|x|^2P_i{}^j\rho_j,
 \qquad R_i^{w^2}=2A_i^{(2)}\Gamma_{11},
 \\
 &\te  U_i^x=\frac1{\sqrt2}e_i{}^\alpha x^p\Gamma_{\alpha p}
 -\frac1{\sqrt2}\partial_ix^p\Gamma_p\Gb,\qquad U_i^{x^2}=-\frac12x^p\partial_ix^q\Gamma_{pq}
 -\frac12|x|^2e_i{}^\alpha\Gamma_\alpha\Gb,\\
&\te
 U_i^w=-\frac{\ii}{2\sqrt2}\Gamma_{11}
 (\Gamma_a^{\rm CP}\partial_iw^a-\bar\Gamma_a^{\rm CP}\partial_i\bar w^a)
 +\frac1{2\sqrt2}
 (\Gamma_a^{\rm CP}\partial_iw^a+\bar\Gamma_a^{\rm CP}\partial_i\bar w^a)\Gb,
 \quad 
 U_i^{w^2}=Q_i^{(2)}+\frac12A_i^{(2)}\Gamma_{11}\Gb.
 \la{233}
\end{align}
If we set $R_i^r=(R_i^r)^A\Gamma_A$, the projector parts  are
\begin{align}
 \Gamma_x^{(1)}={}&\te \frac{\ii}{2\sqrt g}\epsilon^{ijk}
 (R_i^x)^A\rho_j^B\rho_k^C\Gamma_{ABC},\qquad 
 \Gamma_w^{(1)}=\frac{\ii}{2\sqrt g}\epsilon^{ijk}
 (R_i^w)^A\rho_j^B\rho_k^C\Gamma_{ABC},
 \la{235}\\
 \Gamma_{x^2}^{(2)}={}&\te -\frac12H^x\Gk
 +\frac{\ii}{2\sqrt g}\epsilon^{ijk}
 \big[(R_i^{x^2})^A\rho_j^B\rho_k^C
 +(R_i^x)^A(R_j^x)^B\rho_k^C\big]\Gamma_{ABC},\\
 \Gamma_{xw}^{(2)}={}&\te\frac{\ii}{2\sqrt g}\epsilon^{ijk}
 \big[(R_i^x)^A(R_j^w)^B+(R_i^w)^A(R_j^x)^B\big]
 \rho_k^C\Gamma_{ABC},\\
 \Gamma_{w^2}^{(2)}={}&\te-\frac12H^w\Gk
 +\frac{\ii}{2\sqrt g}\epsilon^{ijk}
 \big[(R_i^{w^2})^A\rho_j^B\rho_k^C
 +(R_i^w)^A(R_j^w)^B\rho_k^C\big]\Gamma_{ABC}.
 \la{236}
\end{align}
Then  we get  for the mixed  parts of the quartic Lagrangian
\begin{align}
&\cL_{2b,2f}=\cL_{x^2\theta^2}
 +\cL_{xw\theta^2}+\cL_{w^2\theta^2}, \la{226}\\
 \cL_{x^2\theta^2}={}&\te
 \big(\tfrac14H^xg^{ij}-\tfrac12(H^x)^{ij}\big)
 \tb\rho_iD_j^{(0)}\theta
 +\frac12g^{ij}\tb\big[
 R_i^{x^2}D_j^{(0)}+R_i^xU_j^x+\rho_iU_j^{x^2}
 \big]\theta\no \\
&\te
 -\frac14g^{ij}\tb\Gamma_x^{(1)}
 \big[R_i^xD_j^{(0)}+\rho_iU_j^x\big]\theta
 -\frac14g^{ij}\tb\Gamma_{x^2}^{(2)}\rho_iD_j^{(0)}\theta 
 \la{237}\ , 
\\
 \cL_{xw\theta^2}={}&\te \frac12g^{ij}\tb
 \big[R_i^xU_j^w+R_i^wU_j^x\big]\theta
 -\frac14g^{ij}\tb\Gamma_x^{(1)}
 \big[R_i^wD_j^{(0)}+\rho_iU_j^w\big]\theta\no \\&\te
 -\frac14g^{ij}\tb\Gamma_w^{(1)}
 \big[R_i^xD_j^{(0)}+\rho_iU_j^x\big]\theta
 -\frac14g^{ij}\tb\Gamma_{xw}^{(2)}\rho_iD_j^{(0)}\theta, 
 \la{238}\\
 \cL_{w^2\theta^2}={}&\te
 \big(\frac14H^wg^{ij}-\frac12(H^w)^{ij}\big)
 \tb\rho_iD_j^{(0)}\theta
 +\frac12g^{ij}\tb\big[
 R_i^{w^2}D_j^{(0)}+R_i^wU_j^w+\rho_iU_j^{w^2}
 \big]\theta\no \\ 
 &\te
 -\frac14g^{ij}\tb\Gamma_w^{(1)}
 \big[R_i^wD_j^{(0)}+\rho_iU_j^w\big]\theta
 -\frac14g^{ij}\tb\Gamma_{w^2}^{(2)}\rho_iD_j^{(0)}\theta .
 \la{239}
\end{align}

\section{Four-fermion term}\label{apd}


The exact fermionic part of  $\hat C_3$ in \rf{114}   is generated by  \cite{deWit:1998yu}
\begin{equation}
 \hat C_3=C_3-\int_0^1\dd t\,
 \bar\Theta\Gamma_{AB}E(t\Theta)
 \wedge E^A(t\Theta)\wedge E^B(t\Theta),
 \la{240}
\end{equation}
where $C_3$ is the bosonic 3-form  potential pulled back with
bosonic vielbein legs.
Using \rf{1166}  we have 
\al{
& E(t\Theta)=tD\Theta+\tfrac{1}{6}t^3\cM^2(\Theta)D\Theta+\OO(\Theta^5),
 \qquad
 E^A(t\Theta)=e^A+t^2\bar\Theta\Gamma^AD\Theta+\OO(\Theta^4),\\
&
 \hat C_f^{(2)}=\te -\frac12\bar\Theta\Gamma_{AB}D\Theta
 \wedge e^A\wedge e^B,\no \\
 &\te \hat C_f^{(4)}=-\frac1{24}\bar\Theta\Gamma_{AB}\cM^2(\Theta) D\Theta
 \wedge e^A\wedge e^B
 -\frac12(\bar\Theta\Gamma_{AB}D\Theta)
 \wedge(\bar\Theta\Gamma^AD\Theta)\wedge e^B.
 \no 
}
In terms of  $\theta = 2\sqrt 2 \Theta$ and $\cM^2(\theta) $  this  gives 
\begin{align}\te 
 \hat C_f^{(2)}=-\frac1{16}\tb\Gamma_{AB}D\theta
 \wedge e^A\wedge e^B,\quad 
 \hat C_f^{(4)}=\te -\frac1{1536}\tb\Gamma_{AB}\cM^2(\theta)D\theta
 \wedge e^A\wedge e^B
-\frac1{128}(\tb\Gamma_{AB}D\theta)
 \wedge(\tb\Gamma^AD\theta)\wedge e^B.
 \la{242}
\end{align}
Let us define 
\begin{equation}\la{2411}
 B_{ij}=\tb\rho_iD_j^{(0)}\theta
\ , \qquad \qquad 
 \Pi_{ij}=(\tb\Gamma^AD_i^{(0)}\theta)
           (\tb\Gamma_AD_j^{(0)}\theta).
\end{equation}
From \rf{117}, the fermionic corrections to the induced metric are
\begin{align}
 h_{ij}^{(2f)}=\te \frac1{16}(B_{ij}+B_{ji}),
 \qquad \qquad 
 h_{ij}^{(4f)}=\te \frac1{64}\Pi_{ij}
 +\frac1{1536}\big[
 \tb\rho_i\cM^2D_j^{(0)}\theta+(i\leftrightarrow j)\big].
 \la{245}
\end{align}
The volume  part of the Lagrangian in \rf{114} gives
\begin{align}
 \cL_{4f}^{\rm (vol)}=\te
 \frac1{384}g^{ij}\tb\rho_i(\cM^2)_+D_j^{(0)}\theta
 +\frac1{32}g^{ij}\Pi_{ij}
 -\frac1{32}B_{ij}B^{ij}
 -\frac1{32}B_{ij}B^{ji}
 +\frac1{32}(B_i{}^i)^2.
 \la{246}
\end{align}
Under the $\kappa$-gauge  condition 
$
 g^{ij}\Pi_{ij}=B_{ij}B^{ij}
$, $ \rho^i\Gk=\frac{\ii}{2\sqrt g}
      \epsilon^{ijk}\rho_{jk}$ and thus\foot{Here $\epsilon^{ijk}$ and $\epsilon_{ijk}$ denote numerical alternating
symbols, with $\epsilon^{123}=\epsilon_{123}=1$.  Equivalently, introducing
the Levi--Civita tensors
$\varepsilon^{ijk}={1\ov \sqrt g} \epsilon^{ijk}$ and
$\varepsilon_{ijk}=\sqrt g\,\epsilon_{ijk}$, one has
$\rho_i\Gk={\ii\ov 2}\varepsilon_i{}^{jk}\rho_{jk}$.
For example, in a local $\AdS_2$ orthonormal frame,
$g_{ij}=\operatorname{diag}(1,1,4k^{-2})$, the second identity gives
$\rho_1\Gk=\ii\rho_2\Gamma_{11}$, as required directly from
$\Gk=\ii\rho_1\rho_2\Gamma_{11}$.}  
\al{
 \cL_{4f}^{\rm (vol)}=\te
 \frac1{384}g^{ij}\tb\rho_i(\cM^2)_+D_j^{(0)}\theta
 -\frac1{32}B_{ij}B^{ji}
 +\frac1{32}(B_i{}^i)^2
. 
 \la{248}}
 For the above  choice of the orientation
that produces the projector $(1-\Gamma_{\rm B})$ in the quadratic action, the
quartic Wess--Zumino term  contribution $\cL_{4f}^{\rm (wz)}$    equals that of 
 the  volume part   in \rf{248}.
As a result, 
\begin{equation}
 \cL_{4f}= \cL_{4f}^{\rm (vol)} + \cL_{4f}^{\rm (wz)}=\te    \frac1{192}g^{ij}\tb\rho_i(\cM^2)_+D_j^{(0)}\theta
 -\frac1{16}g^{il}g^{jk}B_{ij}B_{kl}
 +\frac1{16}(g^{ij}B_{ij})^2.
 \la{250}
\end{equation}
All world-volume indices in \rf{250} take the values $1,2,3$, in  
particular,
\begin{align}\te
 \frac1{192}g^{ij}\tb\rho_i(\cM^2)_+D_j^{(0)}\theta
 =\frac1{192}g^{\alpha\beta}
 \tb\rho_\alpha(\cM^2)_+D_\beta^{(0)}\theta
 +\frac{k}{384}\tb\Gamma_{11}(\cM^2)_+
 D_3^{(0)}\theta.
 \la{251}
\end{align}

\section{Propagator normalizations  and  Wick contraction rules}\la{ape}

To recall, with $\s^3\equiv \s^3 + 2 \pi$   
the 33  component of the 3d  metric $g_{ij}$  with unit-radius $\AdS_2$ part  is 
$g_{33}=4/k^2$. The length of the circle is then (cf. \rf{8})
\begin{equation}
 L=\int_0^{2\pi}\dd\sigma^3\sqrt{g_{33}}=\tfrac{4\pi}{k}\ . \la{e1}
\end{equation}
The expansion of 3d fluctuation fields in orthonormal Fourier mode  functions is 
\begin{equation}
 \Phi=\sum_{n=-\infty}^\infty f_n\Phi_n \ , \qquad \ \ 
 f_n(\sigma^3)=\tfrac{1}{ \sqrt L} \,e^{\ii n\sigma^3},
 \qquad
 \int_0^{2\pi}\dd\sigma^3\sqrt{g_{33}}\,f_n f_m
 =\delta_{n+m,0}\ .  \la{141}
\end{equation}
With the convention \rf{141}, the 2d  mode fields in the 
action  in \rf{160}--\rf{18} are canonically normalized.   The kinetic operators and  Green's functions are 
given by 
($\nabla^2\equiv \nabla^2_{\AdS_2}$)\foot{Note that in terms of two real components  $x^p$ of 
  $x=x^1+\ii x^2$, the  presence of the  Wess--Zumino term in the 3d action implies that 
$G_{x,n}\ne G_{x,-n}$  and 
$
 \langle x_n^p\,  x_m^q\rangle
 =\frac{1}{4} \delta_{n+m,0} \big[
 (G_{x,n}+G_{x,-n})\delta^{pq}
 +\ii(G_{x,n}-G_{x,-n})\epsilon^{pq}\big].
$
Only in the charge-symmetric limit  does this reduce to
$\frac12\delta^{pq}G_x$.}
\begin{equation}
 K_{x,n}=-\nabla^2+m_{x,n}^2,
 \qquad K_{w,n}=-\nabla^2+m_{w,n}^2,
 \qquad G_{x,n}=K_{x,n}^{-1},\quad G_{w,n}=K_{w,n}^{-1},\qquad G_{\theta,n}=\mathscr D_n^{-1} \ , \la{e4}
\end{equation}
where     $\mathscr D_n$ in \rf{126},\rf{128}  is  defined on the physical $P_-$ spinor subspace \rf{119}:
\begin{center}
\small
\begin{tabular}{@{}lll@{}}
\toprule
field & quadratic action &  basic contraction\\
\midrule
$x_n$ & $\bar x_{-n}K_{x,n}x_n$
 & $\langle x_n\bar x_m\rangle=\delta_{n+m,0}G_{x,n}$\\
$w_n^a$ & $\bar w^a_{-n}K_{w,n}w_n^a$
 & $\langle w_n^a\bar w_m^b\rangle=\delta^{ab}\delta_{n+m,0}G_{w,n}$\\
$\theta_n$ & $\tfrac12\theta_{-n}^{T}C\mathscr D_n\theta_n$
 & $\langle\theta_n\bar\theta_m\rangle=\delta_{n+m,0}G_{\theta,n}$\\
\bottomrule
\end{tabular}
\end{center}

Let us review  the Majorana fermion  Wick contraction rules 
(see also Appendix C of \cite{Beccaria:2025ahf} and
Appendix A of \cite{Beccaria:2026ffm}). 
We use the formal Euclidean continuation of the Majorana spinor conventions 
$
 \bar\theta=\theta^T C,
 \  C^T=-C,
 \  C^{-1}\Gamma_A^T C=-\Gamma_A.
$
For two identical Grassmann spinors only the antisymmetric part of the matrix
multiplying $\theta^T\theta$ contributes  so that for a local bilinear
$\bar\theta W\theta$  we have 
\be 
 CW \longrightarrow \tfrac12\big[CW-(CW)^T\big],
\qquad  
 W\to  W_{\rm A}
 \equiv\tfrac12\big(W+C^{-1}W^TC\big) \ , \quad (CW_{\rm A})^T=-CW_{\rm A} \ , \quad  C^{-1}W_{\rm A}^TC=W_{\rm A}\ .   \ee
 For a  Majorana fermion with an 
antisymmetric kinetic operator  $\KK=C\mathscr D$ we have 
\begin{equation}
 \int\dd\theta\,e^{-\frac12\theta^T\KK\theta+\eta^T\theta}
 =\operatorname{Pf}(\KK)\  e^{-\frac12\eta^T\KK^{-1}\eta}\ ,\qquad \ \ 
 \qquad \langle\theta\, \theta^T\rangle=\KK^{-1}\ . \la{e5}
\end{equation}
For 3d   fermion field
\begin{equation}
 G_\theta(\sigma,\sigma')
 \equiv\big\langle\theta(\sigma)\, \bar\theta(\sigma')\big\rangle, \qquad 
 \big\langle\theta_\a(\sigma)\theta_\b(\sigma')\big\rangle
 =\big[G_\theta(\sigma,\sigma')C^{-1}\big]_{\a\b},
 \qquad
 \big[G_\theta(\sigma,\sigma')C^{-1}\big]^T
 =-G_\theta(\sigma',\sigma)C^{-1}.
 \la{147}
\end{equation}
In the  coincident  point case for  projected matrices $W$ and $V$ 
(here the trace is over the physical 16-component spinor space)
\begin{equation}
 \big\langle\bar\theta W\theta\big\rangle
 =-\tr\big(G_\theta W\big)\ , \qquad \qquad \langle\tb W\theta\ \tb V\theta\rangle
 =\tr(WG_\theta)\ \tr(VG_\theta)
 -2\tr(W\, G_\theta\,  V\, G_\theta)\ . 
 \la{148}
 \end{equation}
Let $\mathscr D_i$ denote any derivative 
(e.g.  $\nabla_a$, $D_a^{(0)}$, $\partial_3$ or
$D_3^{(0)}$)  acting on the unbarred spinor. 
Then let us define  $ G_{\theta,i}$    and  $G_{\theta,ij}$ such that  
\begin{align}
 G_{\theta,i}
 &\equiv
 \big\langle(\mathscr D_i\theta)\bar\theta\big\rangle,
 \qquad  \qquad 
G_{\theta,ij}C^{-1}
 \equiv
 \big\langle(\mathscr D_i\theta)
 (\mathscr D_j\theta)^T\big\rangle.
 \la{150}
\\
 \big\langle\bar\theta W\mathscr D_i\theta\big\rangle
 &=-\tr\big(G_{\theta,i}W\big),
 \qquad\qquad 
 \big\langle(\mathscr D_i\theta)_\a\theta_\b\big\rangle
 =\big(G_{\theta,i}C^{-1}\big)_{\a\b}.
 \la{151}
\end{align}
 With one differentiated unbarred
spinor  and  $V^c=C^{-1}V^TC$  we have 
\begin{align}
 &\big\langle
 (\bar\theta W\mathscr D_i\theta)
 (\bar\theta V\theta)
 \big\rangle
=\tr(WG_{\theta,i})\tr(V G_\theta)
 -\tr\, \big[WG_{\theta,i}
 (V+V^c)G_\theta\big],
 \la{153}\\
  &\big\langle
 (\bar\theta W\mathscr D_i\theta)
 (\bar\theta V\mathscr D_j\theta)
 \big\rangle
 =
 \tr(WG_{\theta,i})\tr(V G_{\theta,j})
 -\tr(WG_{\theta,i}V G_{\theta,j})
 -\tr\, \big(
 WG_{\theta,ij}C^{-1}V^TCG_\theta
 \big).
 \la{154}
\end{align}

Let  us  now discuss  the  decomposition of  the fermion propagator  at coincident points 
into  KK tower subsectors defined in Appendix \ref{apb},
\begin{equation}
 G_\theta(\sigma,\sigma)=L^{-1}
 \sum_{n=-\infty}^\infty\sum_\T G_{\theta;\T,n}\  , \qquad \qquad \big\langle\theta_n\bar\theta_m\big\rangle
 =\delta_{n+m,0}\sum_\T G_{\theta;\T,n}.
 \la{166}
\end{equation}
In discussing coincident  limit of the propagators   we always assume the 
 use of the analytic
regularization  prescription  in which  $\delta(0)=0$.
$\G(m)$ will denote the regularized scalar coefficient of the coincident
fermion Green's function.

 We  have  (we fix one fermionic  set  $\T$ and one mode number $n$)
\begin{equation}
 \bar\theta_{\T,-n}=\bar\theta_{-n}\Pi_\T
   =(\theta_{\T^c,-n})^TC,
 \qquad\qquad \ \ 
 G_{\theta;\T,n}=-\Gb\Pi_\T \G_{\T,n},
 \qquad \G_{\T,n}=\G(m_{\T,n}),
 \qquad \tr\,\Pi_\T=2.
 \la{156}
\end{equation}
For the correlators   with $\AdS_2$-derivatives in  \rf{150}  we have ($m_{\T^c,-n}=m_{\T,n}$)
\al{
 &G_{\theta,a;\T,n}\equiv\big\langle\nabla_a\theta_{\T,n}
 \bar\theta_{\T,-n}\big\rangle
 =-\ha {m_{\T,n}}\Gamma_a\Pi_\T \G_{\T,n}\ , \la{157}\\
 &
  U_{ab;\T,n}
\equiv
 \big\langle\nabla_a\nabla_b\theta_{\T,n}
 \bar\theta_{\T,-n}\big\rangle=
 \tfrac14\big[(2m_{\T,n}^2-1)g_{ab}-\Gamma_{ab}\big]
 G_{\theta;\T,n}\ , 
 \la{158}
\\
 &G_{\theta,ab;\T,n}C^{-1}
 \equiv
 \big\langle(\nabla_a\theta_{\T,n})
 (\nabla_b\theta_{\T^c,-n})^T\big\rangle,
 \nonumber\\
 &G_{\theta,ab;\T,n}
 =-\tfrac14\big[
 \Gamma_{ab}+(2m_{\T,n}^2-1)g_{ab}
 \big]G_{\theta;\T,n}
=\tfrac14\big[
 \Gamma_{ab}+(2m_{\T,n}^2-1)g_{ab}
 \big]\Gb\Pi_\T \G_{\T,n}.
 \la{159}
\\
&\te
 g^{ab}G_{\theta,ab;\T,n}
 =-\big(m_{\T,n}^2-\frac12\big)G_{\theta;\T,n},
 \qquad
 G_{\theta,ab;\T,n}
 -G_{\theta,ba;\T,n}
 =-\frac12\Gamma_{ab}G_{\theta;\T,n}.
 \la{1601}
}
Note also that since 
$ (\Gamma^a\nabla_a)^2=\nabla^2-\frac{1}{4} R
 =\nabla^2+\frac12,$   ($R(\AdS_2)=-2$)
we get,    using $\delta(0)=0$, 
$
 g^{ab}U_{ab;\T,n}
=\big(m_{\T,n}^2-\tfrac12\big)G_{\theta;\T,n}.
$
The antisymmetric part of \rf{158}
 follows independently from
$
 [\nabla_a,\nabla_b]=\frac14R_{ab}{}^{cd}\Gamma_{cd}
 =-\frac12\Gamma_{ab},
$ i.e.  $
 U_{ab;\T,n}-U_{ba;\T,n}=-\frac12\Gamma_{ab}G_{\theta;\T,n}.$
 
Synge's rule, i.e. $\nabla_a[F]=[\nabla_aF]+[\nabla_{a'}F]$ for the coincidence
limit $[F]$ of any bitensor $F(\s,\s')$, 
applied to 
$\langle\nabla_a\theta(x)\bar\theta(x')\rangle$ gives the exact relation
\begin{equation}
 G_{\theta,ab;\T,n}=- U_{ba;\T,n}.
 \la{161}
\end{equation}
Similarly, for the derivative that appears in the fermionic  action,
$D_a^{(0)}=\nabla_a-\frac12\Gamma_a\Gb$, one finds
\al{
&\qquad \qquad  G^{(D)}_{\theta,a;\T,n}
 = -\ha (m_{\T,n}+1)\Gamma_a\Pi_\T \G_{\T,n},\\
&
 G^{(D)}_{\theta,ab;\T,n}
 =-\ha (m_{\T,n}+1)
 \big(\Gamma_{ab}+m_{\T,n}g_{ab}\big)G_{\theta;\T,n}
=\ha ({m_{\T,n}+1})
 \big(\Gamma_{ab}+m_{\T,n}g_{ab}\big)
 \Gb\Pi_\T \G_{\T,n}.
 \la{163}
}
For $\del_3$  and $D_3$ derivative insertions we have 
\al{
& G^{(\partial_3)}_{\theta;\T,n}=\ii nG_{\theta;\T,n},
 \qquad
 G^{(\partial_3\partial_3)}_{\theta;\T,n}
 =n^2G_{\theta;\T,n},
 \qquad
 \big\langle\partial_3^2\theta_{\T,n}
 \bar\theta_{\T,-n}\big\rangle=-n^2G_{\theta;\T,n}.
 \la{164}
\\
& G^{(D_3)}_{\theta;\T,n}=\ii q_{\T,n}G_{\theta;\T,n},
 \qquad
 G^{(D_3D_3)}_{\theta;\T,n}=q_{\T,n}^2G_{\theta;\T,n}.
 \la{165}
}
Here $\ii  q_{\T,n}$ is the eigenvalue of $D_3^{(0)}$ in \rf{124} 
on the $(\T,n)$ block, i.e. 
$q_{\T,n}= n+\tfrac1k\big(\xi_\T+\tfrac12\epsilon_\T\big)
=\tfrac{2}{k}\,\epsilon_\T\big(m_{\T,n}+1\big)$.




\section{Expectation value of  quartic bosonic vertex}
\label{apf}

Here we   discuss  the computation of  the expectation  value \rf{55}  of $ \L_{4b}  $  in \rf{52}. 
  We assume the analytic regularization 
prescription  in which  $\delta(0)=0$. 

We shall use the following notation for the  scalar   Green's function values at the coincident points:
\begin{align}
 \langle x\bar x\rangle&=\G_x, \qquad \ \   \langle\partial_3x\,\bar x\rangle=\ii \G_x^{(3)}, \qquad \ \ 
 \langle\partial_3x\,\partial_3\bar x\rangle=\G_x^{(33)},\qquad  \ \
  \langle\partial_\alpha x\,\partial_\beta\bar x\rangle
 =g_{\alpha\beta}\G_x^{(dd)},  \la{194}\\
 \langle w^a\bar w^b\rangle&=\delta^{ab}\G_w,
 \qquad 
\langle\partial_3w^a\,\bar w^b\rangle =\ii\delta^{ab}\G_w^{(3)},
 \qquad 
 \langle\partial_3w^a\,\partial_3\bar w^b\rangle
 =\delta^{ab}\G_w^{(33)},
\qquad  \langle\partial_\alpha w^a\,\partial_\beta\bar w^b\rangle
 =g_{\alpha\beta}\delta^{ab}\G_w^{(dd)},\no
\end{align}
and similarly    $ \langle x\,\partial_3\bar x\rangle=-\ii \G_x^{(3)},
 \ \ \langle\bar x\,\partial_3x\rangle=\ii \G_x^{(3)},$ etc.
 Note that $ \langle x\,\partial_a\bar x\rangle= 0$  due to $\AdS_2$ covariance
 and for $J_3$ in \rf{45} we have 
  $\langle J_3\rangle
 =\frac{\ii k}{4}
 \big(\langle w^a\partial_3\bar w^a\rangle
       -\langle\bar w^a\partial_3w^a\rangle\big)
 =\frac{3k}{2}\G_w^{(3)}$.

 In terms of the
canonically normalized 2d  mode propagators 
\begin{align}\la{1990}
 \G_x =L^{-1}\sum_n\G_{x,n},
 \quad  \G_x^{(3)} =L^{-1}\sum_n n\G_{x,n},
 \quad 
 \G_x^{(33)}=L^{-1}\sum_n n^2\G_{x,n}, \quad 
 \G_x^{(dd)}=L^{-1}\sum_n\G_{x,n}^{(dd)}, \   {\rm etc.} \end{align}

The quadratic bosonic Lagrangian following from \rf{46} and
\rf{48} is given in \rf{17}, i.e. 
\begin{align}
 \cL_{2b}=&\te 
 g^{\alpha\beta}\nabla_\alpha x\partial_\beta\bar x
 +\frac{k^2}{4}\partial_3x\partial_3\bar x
 +2x\bar x
 -\frac{3\ii k}{4}
 \big(x\partial_3\bar x-\bar x\partial_3x\big)\no \\
 &\te \la{2000}
 +g^{\alpha\beta}\partial_\alpha w^a\partial_\beta\bar w^a
 +\frac{k^2}{4}\partial_3w^a\partial_3\bar w^a
 +\frac{\ii k}{4}
 \big(w^a\partial_3\bar w^a-\bar w^a\partial_3w^a\big).
\end{align}
The corresponding kinetic operators are 
\begin{align}\te 
 \te K_x=-\nabla^2-\frac{k^2}{4}\partial_3^2
       +\frac{3\ii k}{2}\partial_3+2\ ,
 \qquad \ \ 
 K_w=-\nabla^2-\frac{k^2}{4}\partial_3^2
       -\frac{\ii k}{2}\partial_3\ , 
 \la{202}
\end{align}
and 
$\langle(K_xx)\bar x\rangle=0$ and
$\langle(K_ww^a)\bar w^b\rangle=0$ imply the ``equations of motion'' relations 
\begin{align}
\te  2\G_x^{(dd)}+\frac{k^2}{4}\G_x^{(33)}
 -\frac{3k}{2}\G_x^{(3)}+2\G_x=0,
 \la{203}\qquad \ \ 
 2\G_w^{(dd)}+\frac{k^2}{4}\G_w^{(33)}
 +\frac{k}{2}\G_w^{(3)}=0.
\end{align}
Doing Wick contractions, from \rf{49} we find 
\begin{align}
\te  \langle\cL_{4x}\rangle=
 4\G_x^2+2\G_x\G_x^{(dd)}-2\big(\G_x^{(dd)}\big)^2
+\frac{3k^2}{4}
 \big[\G_x\G_x^{(33)}+\big(\G_x^{(3)}\big)^2\big]
 -\frac{k^4}{16}\big(\G_x^{(33)}\big)^2.
 \la{210}
\end{align}
Similarly, from \rf{50}  and \rf{51} we get 
\begin{align}
 \langle\cL_{2x,2w}\rangle=&\te 
 \G_x\big(3k\G_w^{(3)}+\frac{3k^2}{2}\G_w^{(33)}\big)
 +\G_x^{(33)}\big(
 \frac{3k^2}{2}\G_w^{(dd)}
 -\frac{3k^4}{16}\G_w^{(33)}
 -\frac{3k^3}{8}\G_w^{(3)}\big)\no \\
 &\te +\G_x^{(dd)}\big(
 \frac{3k^2}{2}\G_w^{(33)}+3k\G_w^{(3)}\big),
 \la{215}
\\
 \langle\cL_{4w}\rangle={}&\te
 -\frac{3k^2}{2}\big(\G_w^{(3)}\big)^2
 -\frac{3k^3}{2}\G_w^{(3)}\G_w^{(33)}
 -\frac{3k^4}{8}\big(\G_w^{(33)}\big)^2
 \nonumber\\
 &\te -6\big(\G_w^{(dd)}\big)^2
 -3k\G_w\G_w^{(3)}-\frac{3k^2}{2}\G_w\G_w^{(33)}
 -12\G_w\G_w^{(dd)}
 +6k\G_w^{(3)}\G_w^{(dd)}
 +3k^2\G_w^{(dd)}\G_w^{(33)}.
 \la{219}
\end{align}
The  quartic  part of the bosonic Wess--Zumino term in \rf{48}   gives 
\al{ \cL_{4b}^{\rm (wz)}
 \te =-\frac{9\ii k}{8}|x|^2
 \big(x\partial_3\bar x-\bar x\partial_3x\big)\ , \ \ \ \qquad \ \
\te  \langle\cL_{4b}^{\rm WZ}\rangle
 =-\frac{9k}{2}\G_x\G_x^{(3)}.
 \la{221}
}
Using \rf{203} to eliminate   $\G_x^{(dd)}$  and $\G_w^{(dd)}$ 
we find that terms with $\G_x$ and $\G_w$ cancel out   and we finish with the ``perfect 
square''  expression in \rf{55}, i.e. 
\al{
 &\langle\cL_{4b}\rangle
 =\langle\cL_{4x}\rangle+
  \langle\cL_{2x,2w}\rangle+
  \langle\cL_{4w}\rangle+
  \langle\cL_{4b}^{\rm (wz)}\rangle \te = -\tfrac{3}{8}\cX^2_b \ ,\la{2666} \\
  &\cX_b = -\tfrac{k}{2}
 \Big[ 2\G_x^{(3)}-k\G_x^{(33)}
 -3\big(2\G_w^{(3)}+k\G_w^{(33)}\big)\Big]\ . 
 \la{224}
}

\section{Expectation value of  mixed  boson--fermion quartic  vertex}
\label{apg}

Here we  compute the expectation value of the mixed quartic term in the Lagrangian in \rf{56} or \rf{226}--\rf{239}. 

Let us start with introducing some notation. We shall use  a local orthonormal world-volume frame with indices 
$\widehat i=(a,\widehat{3})$, where $a=1,2$, and set
\begin{equation}
 \rho_a=\Gamma_a,\qquad\ \ 
 \rho_{\widehat{3}}=\Gamma_{11},\qquad \del _{\widehat{3}}=\tfrac{k}{2}\del_3 \ . 
  \la{268}
\end{equation}
For  fixed  fermion set  and Fourier mode we define (cf. \rf{157},\rf{135})
\begin{equation}
 r_{\T,n}\equiv m_{\T,n}+1 \ . 
 \la{269}
\end{equation}
As in Appendix \ref{ape}   we have for coincident fermionic  Green's functions 
\begin{align}
&\la{g3}
 G_{\T,n} \equiv \G_{\theta;\T,n}
   =-\Gb\Pi_\T \G_{\T,n},
\qquad \ \  \ G^{(D)}_{a;\T,n}
   =-\tfrac{1}2r_{\T,n}\Gamma_a\Pi_\T \G_{\T,n},
 \qquad  G^{(D)}_{\widehat{3};\T,n}
   =r_{\T,n}\Gamma_{11}\Pi_\T \G_{\T,n}, 
 \\ &
\te  G^{(D)}_{ab;\T,n}
 =\tfrac{1}2r_{\T,n}
   \big(\Gamma_{ab}+m_{\T,n}g_{ab}\big)
   \Gb\Pi_\T \G_{\T,n},
 \qquad \qquad 
 G^{(D)}_{a\widehat{3};\T,n}
 =G^{(D)}_{\widehat{3}a;\T,n}
 =\frac{\ii}{2} \epsilon_\T r_{\T,n}^2
   \Gamma_a\Pi_\T \G_{\T,n},\no \\
&\qquad  \qquad
 G^{(D)}_{\widehat{3}\widehat{3};\T,n}
 =r_{\T,n}^2  G_{\T,n}
 =-r_{\T,n}^2\Gb\Pi_\T \G_{\T,n}.\la{g44}
\end{align}
It is convenient to introduce also the  following notation for the sums over $n$ (here  $\G_{\T,n} \equiv \G_f(m_{\T,n})$ in  \rf{60})
\begin{align}
&  \GG \equiv \sum_{\T} \GG_\T, \qquad \qquad 
\GG_\T\equiv
 L^{-1}\sum_{n=-\infty}^\infty \G_{\T,n},
 \qquad \qquad \GG_{\rm tr}\equiv\sum_\T\chi_\T\GG_\T
 =\sum_{\T:\,|\xi_\T|={1\ov 2}}\GG_\T, 
 \la{277}\\  &   \cX_f \equiv \G^{(1)} \equiv  \sum_\T   \G^{(1)}_\T,  \qquad \ \ 
 \G^{(1)}_\T\equiv
 L^{-1}\sum_{n=-\infty}^\infty
 r_{\T,n}\G_{\T,n},\qquad  
 \qquad 
 \G^{(2)}_\T\equiv
 L^{-1}\sum_{n=-\infty}^\infty
 r_{\T,n}^2\G_{\T,n}. \la{279}
\end{align}
$\hat \G_{\rm tr}$ is the triplet-projected sum where $\chi_\T$ 
is the  weight that selects the 6 triplet components   of 2d fermions 
($\xi_\T\epsilon_\T=-\frac12$ on 6 triplet components  and
$\xi_\T\epsilon_\T=\frac32$ on 2 singlet ones)
\begin{equation}
 \chi_\T\equiv\tfrac14\big(3-2\xi_\T\epsilon_\T\big)
 =\begin{cases}
  1, & |\xi_\T|=\frac12,\\[1mm]
  0, & |\xi_\T|=\frac32.
 \end{cases}\la{2801}
\end{equation}
We have 
\begin{align}
 \big\langle\tb\rho_{\widehat i}
 D_{\widehat j}^{(0)}\theta\big\rangle
 =\G^{(1)}\,
 \operatorname{diag}(1,1,-2)_{\widehat i\widehat j}\ ,
 \qquad \qquad 
 \big\langle\tb M\theta\big\rangle
 =\sum_\T\G_\T\,
 \tr\, \big(\Gb\Pi_\T M\big)\ , 
 \la{256}
\end{align}
where $M$ is any  spinor matrix.
  In particular,
the trace term  in \rf{226} has vanishing  expectation value  after use of the   fermion equation of motion: 
\be 
 \big\langle H g^{ij}\tb\rho_iD_j^{(0)}\theta\big\rangle
 =\langle H\rangle
 \sum_{\widehat i}
 \big\langle\tb\rho_{\widehat i}
 D_{\widehat i}^{(0)}\theta\big\rangle=0.
\ee
The bosonic metric insertions have the diagonal expectations
\begin{align}
 \big\langle H^x_{\widehat i\widehat j}\big\rangle
 &=2\operatorname{diag}\!\big(
 \G_x^{(dd)}+\G_x,\,
 \G_x^{(dd)}+\G_x,\,\tfrac{k^2}{4} \G^{(33)}_x\big),
 \qquad 
 \big\langle H^w_{\widehat i\widehat j}\big\rangle
 =6\operatorname{diag}\!\big(
 \G_w^{(dd)},\,\G_w^{(dd)},\,
 \tfrac{k^2}{4} \G^{(33)}_w+\tfrac{k}{2} \G^{(3)}_w\big).
 \la{258}
\end{align}
The term containing $\Gamma_{(2)}$ in \rf{193} 
 vanishes  due to the 
cyclicity of the trace  and the coincident limit of the Dirac equation
$\sum_{\widehat i}\rho_{\widehat i}\G_{\widehat i}^{(D)}=0$
\begin{equation}
 \sum_{\widehat i}
 \big\langle\tb \Gamma_{(2)}\ 
 \rho_{\widehat i}D_{\widehat i}^{(0)}\theta\big\rangle 
 =-\tr\, \big[
 \big(\sum_{\widehat i}\rho_{\widehat i}
 \G_{\widehat i}^{(D)}\big)
 \langle\Gamma_{(2)}\rangle\big]=0.
 \la{259}
\end{equation}
Note that since the quadratic  action in \rf{17}  has no $x$--$w$ term and thus there is no such entry in the propagator 
the expectation value of \rf{238} is trivial 
\begin{equation}
 {\big\langle\cL_{xw\theta^2}\big\rangle=0.}
 \la{260}
\end{equation}
Table~\ref{tabx} lists contributions of  the seven structures in $\L_{x^2\theta^2}$ in 
\rf{237}.  The
second and third columns give the coefficients multiplying $\G^{(1)}$ and
$\hat \G$ defined  in \rf{279},\rf{277}, respectively.
\begin{table}[H]
\centering
\begin{tabular}{@{}lcc@{}}
\toprule
term in $\cL_{x^2\theta^2}$
 & coefficient of $\G^{(1)}$
 & coefficient of $\hat \G$\\
\midrule
$H^x$  term
 & $2\tfrac{k^2}{4} \G^{(33)}_x-2\G_x^{(dd)}-2\G_x$ & $0$\\
$\tfrac12 R^{x^2}D^{(0)}$
 & $\G_x$ & $0$\\
$\tfrac12 R^xU^x$
 & $0$ & $2\G_x^{(dd)}+\tfrac{k^2}{4} \G^{(33)}_x$\\
$\tfrac12\rho U^{x^2}$
 & $0$ & $\G_x-\tfrac12\tfrac{k}{2} \G^{(3)}_x$\\
$-\tfrac14\Gamma_x^{(1)}R^xD^{(0)}$
 & $\G_x^{(dd)}-\tfrac{k^2}{4} \G^{(33)}_x$ & $0$\\
$-\tfrac14\Gamma_x^{(1)}\rho U^x$
 & $0$ & $-\G_x^{(dd)}-\tfrac12\tfrac{k^2}{4} \G^{(33)}_x-\tfrac{k}{2} \G^{(3)}_x$\\
$-\tfrac14\Gamma_{x^2}^{(2)}\rho D^{(0)}$
 & $0$ & $0$\\
\bottomrule
\end{tabular}
\caption{Contributions to  $\langle \L_{x^2\theta^2}\rangle$}
\label{tabx}
\end{table}
As a result, 
\begin{align}
 \big\langle\cL_{x^2\theta^2}\big\rangle
 ={}&\big(\tfrac{k^2}{4} \G^{(33)}_x-\G_x^{(dd)}-\G_x\big)\cX_f\te 
 +\big(\G_x+\G_x^{(dd)}+\tfrac{k^2}{8} \G^{(33)}_x
 - \tfrac{3k}{4} \G^{(3)}_x\big)\GG.
 \la{261}
\end{align}
The coefficient of $\GG$ vanishes  due to the ``equation of motion'' relation \rf{203} and  solving it for $\G_x^{(dd)}$   we finally get 
\begin{equation}
 \big\langle\cL_{x^2\theta^2}\big\rangle
 =\tfrac{3k}{8}
 \big(k\G_x^{(33)}-2\G_x^{(3)}\big)\cX_f \ . 
 \la{262}
\end{equation}
In the case of $\L_{w^2\theta^2}$   the corresponding 
contributions of various terms in \rf{239}  
are shown in
Table~\ref{tabw}.
\begin{table}[H]
\centering
\begin{tabular}{@{}lcc@{}}
\toprule
term  in $\cL_{w^2\theta^2}$
 & coefficient of $\G^{(1)}$
 & coefficient of $\hat \G_{\rm tr}$\\
\midrule
$H^w$																	 term
 & $6(\tfrac{k^2}{4} \G^{(33)}_w+\tfrac{k}{2} \G^{(3)}_w-\G_w^{(dd)})$ & $0$\\
$\tfrac12 R^{w^2}D^{(0)}$
 & $-3\tfrac{k}{2} \G^{(3)}_w$ & $0$\\
$\tfrac12 R^wU^w$
 & $0$ & $-2(2 \G_w^{(dd)}+\tfrac{k^2}{4} \G^{(33)}_w)$\\
$\tfrac12\rho U^{w^2}$
 & $0$ & $-\tfrac{k}{2} \G^{(3)}_w$\\
$-\tfrac14\Gamma_w^{(1)}R^wD^{(0)}$
 & $3(\G_w^{(dd)}-\tfrac{k^2}{4} \G^{(33)}_w)$ & $0$\\
$-\tfrac14\Gamma_w^{(1)}\rho U^w$
 & $0$ & $2\G_w^{(dd)}+\tfrac{k^2}{4} \G^{(33)}_w$\\
$-\tfrac14\Gamma_{w^2}^{(2)}\rho D^{(0)}$
 & $0$ & $0$\\
\bottomrule
\end{tabular}
\caption{Contributions to  $\langle \L_{w^2\theta^2}\rangle$}
\label{tabw}
\end{table}
For fixed fermion projection $\T$ different terms  contribute as 
\begin{align}
 &\te \frac12R^wU^w:
 \quad \frac{1}{2}\mathsf W_w
 (2\xi_\T\epsilon_\T-3)\G_\T
 =-2\mathsf W_w\chi_\T\G_\T; \qquad 
 \frac12\rho U^{w^2}:
 \te \quad \frac{1}{4}\tfrac{k}{2} \G^{(3)}_w
 (2\xi_\T\epsilon_\T-3)\G_\T
 =-\tfrac{k}{2} \G^{(3)}_w\chi_\T\G_\T;\no \\
 &\te -\frac14\Gamma_w^{(1)}\rho U^w:
 \quad -\frac{1}{4}\mathsf W_w
 (2\xi_\T\epsilon_\T-3)\G_\T
 =\mathsf W_w\chi_\T\G_\T, \qquad \qquad \mathsf W_w\equiv 2\G_w^{(dd)}+\tfrac{k^2}{4} \G^{(33)}_w.
\end{align}
Summing up  all the contributions  gives 
\begin{align}
 \big\langle\cL_{w^2\theta^2}\big\rangle
 ={}&3\big(\tfrac{k^2}{4} \G^{(33)}_w+\tfrac{k}{2} \G^{(3)}_w-\G_w^{(dd)}\big)\cX_f
-\big(2\G_w^{(dd)}+\tfrac{k^2}{4} \G^{(33)}_w+\tfrac{k}{2} \G^{(3)}_w\big)
 \GG_{\rm tr}.
 \la{263}
\end{align}
The second  bracket vanishes   due to \rf{203}
so that we get 
\begin{equation}
 \big\langle\cL_{w^2\theta^2}\big\rangle
  =\tfrac{9k}{8}
 \big(k\G_w^{(33)}+2\G_w^{(3)}\big)\cX_f \ .   \la{264}
\end{equation}
Combining \rf{260}, \rf{262} and
\rf{264} we finish with the expression quoted in \rf{57}
\begin{equation}
 \big\langle\cL_{2b,2f}\big\rangle
 =\tfrac{3k}{8}\big[
 k\G_x^{(33)}-2\G_x^{(3)}
 +3\big(2\G_w^{(3)}+k\G_w^{(33)}\big)\big]\cX_f= \tfrac{3}{4} \cX_b \cX_f \ , 
 \la{266}
\end{equation}
where $\cX_b$ was  defined in \rf{55} or \rf{224}. 


\section{Expectation value of  four-fermion vertex}\label{aph}

We next   evaluate the  expectation value of  $\L_{4f}$ in \rf{59},\rf{250}. 

Using that  $\tr\big(\rho_{\widehat i}G^{(D)}_{\widehat j;\T}\big)
 =\G^{(1)}_\T\,
 \operatorname{diag}(-1,-1,2)_{\widehat i\widehat j}$
we find for the expectation value  of the 
 last two terms in  \rf{250} 
\begin{equation}
 \big\langle \tfrac1{16}
 \big(g^{ij}g^{kl}-g^{il}g^{jk}\big)B_{ij}B_{kl}\big\rangle
 =-\tfrac38(\cX_f)^2
  -\tfrac14\sum_\T\GG_\T\G^{(1)}_\T , \ \ \ \ \ \ \ 
 \la{285}
\end{equation}
where we used  \rf{154}.
All terms proportional to $\G^{(2)}_\T$ cancelled between the $ab$,
$a\widehat{3}$, $\widehat{3}a$ and $\widehat{3}\widehat{3}$  contributions.


Expanding  $(\cM^2)_+$ as in \rf{121}   let us define $K_{\T \T'}$ by
\begin{equation}
 \big\langle
 g^{ij}\tb\rho_i(\cM^2)_+D_j^{(0)}\theta
 \big\rangle
 =\sum_{\T,\T'}\GG_\T\G^{(1)}_{\T'} K_{\T \T'}.
 \la{286}
\end{equation}
For a dyad $X\otimes Y$ contributed by $(\cM^2)_+$, let
$V_{\widehat i}\equiv\rho_{\widehat i}X$ denote the full Clifford matrix
standing between $\tb$ and $\theta$ in the undifferentiated bilinear.  Thus
$V_{\widehat i}$ includes the external $\rho_{\widehat i}$ from
$g^{ij}\tb\rho_i(\cM^2)_+D_j^{(0)}\theta$.
  Let 
$V_{\widehat i}^{c}\equiv C^{-1}V_{\widehat i}^{T}C$.  Then the  Majorana Wick
contraction  gives ($\GG=\G_\theta$)
\begin{align}
 &\big\langle(\tb V_{\widehat i}\theta)
 (\tb YD_{\widehat i}^{(0)}\theta)\big\rangle
 =
 \tr(V_{\widehat i}\G_\theta)
 \tr\big(Y\G_{\widehat i}^{(D)}\big)
 -\tr\, \big[(V_{\widehat i}+V_{\widehat i}^{c})\G_\theta
 Y\G_{\widehat i}^{(D)}\big].
 \la{287}
\end{align}
Equivalently, the undifferentiated bilinear is automatically projected by
$V_{\widehat i}\to\tfrac12(V_{\widehat i}+V_{\widehat i}^{c})$.

  The required physical
16-component Clifford traces are summarized in
Table~\ref{tabt}. Here $\alpha,\beta=1,2$, $p,q=3,4$ and
$I,J=5,\ldots,10$.  In the first column, $X\otimes Y$ displays only the dyad
coming from $(\cM^2)_+$ and separates its undifferentiated and differentiated
factors.  Every table entry already includes the external
$\rho_{\widehat i}$ through $V_{\widehat i}=\rho_{\widehat i}X$, the
charge-conjugation projection in \rf{287}, the
world-volume contraction, and the  sum over the 
target-space indices.
\begin{table}[H]
\centering
\begin{tabular}{@{}lc@{}}
\toprule
sector of $(\cM^2)_+$ & contribution to $K_{\T \T'}$\\
\midrule
$2\Gb\Gamma_\alpha\otimes\Gamma^\alpha$
 & $8+8\delta_{\T \T'}$\\
$\Gb\Gamma_{11}\otimes\Gamma^{11}$
 & $4$\\
$\Gamma_{\alpha\beta}\otimes\Gamma^{\alpha\beta}\Gb$
 & $8\delta_{\T \T'}$\\
$\Gamma_{pq}\otimes\Gamma^{pq}\Gb$
 & $-8$\\
$-\tfrac12\Gamma_{IJ}\otimes\Gamma^{IJ}\Gb$
 &\ \ \ \ \   $-4+16\delta_{\T \T'}+16\delta_{\T',\T^c}$\\
\bottomrule
\end{tabular}
\caption{Spinor trace coefficients in the expectation value of the
$(\cM^2)_+$ term}
\label{tabt}
\end{table}
The term proportional to $\delta_{\T',\T^c}$ comes entirely from the
$\CP^3$ bivector sum (this is a manifestation of the fact that
the corresponding Clifford matrices reverse all three charge-projector
labels).  As a result, $K_{\T \T'}=32\delta_{\T \T'}+16\delta_{\T',\T^c}$ and thus 
\al{ 
  \big\langle
 \tfrac1{192}g^{ij}\tb\rho_i(\cM^2)_+D_j^{(0)}\theta
 \big\rangle
 =\tfrac16\sum_\T\GG_\T\G^{(1)}_\T
  +\tfrac1{12}\sum_\T\GG_\T\G^{(1)}_{\T^c}.
 \la{289}
}
Combining \rf{285} and \rf{289} gives the  pure square expression in \rf{61}
\begin{equation}
 \big\langle\cL_{4f}\big\rangle
 =-\tfrac38\cX_f^2
 +\tfrac1{12}\sum_\T\GG_\T
 \big(\G^{(1)}_{\T^c}-\G^{(1)}_\T\big) = -\tfrac38\cX_f^2 \ . 
 \la{290}
\end{equation}
Here we used that since 
$
 m_{\T^c,-n}=m_{\T,n}$ (or 
 $r_{\T^c,-n}=r_{\T,n}$)
 one has   $ \GG_{\T^c,-n}=\GG_{\T,n}$ and thus 
 after summation over $n$ 
 \be 
 \GG_{\T^c}=\GG_\T,
 \qquad\qquad 
 \G^{(1)}_{\T^c}=\G^{(1)}_\T.
 \la{292}
\end{equation}

\iffa 
The $\AdS_4$ coframe and connection used above are
\begin{align}
 E^\alpha&=\frac12\frac{1+{1\ov 4} X^2}{1-{1\ov 4} X^2}e^\alpha,
 &E^p&=\frac12\frac{\dd X^p}{1-{1\ov 4} X^2},\\
 \Omega^{\alpha p}&=\frac{X^pe^\alpha}{1-{1\ov 4} X^2},
 &\Omega^{pq}&=-\frac{X^p\dd X^q-X^q\dd X^p}
 {2(1-{1\ov 4} X^2)}.
\end{align}
For the Hopf fibration,
\begin{equation}
 \eta=\frac1k\dd y+A,\qquad
 \Omega^a{}_b=\Omega^a{}_b\big|_{\CP^3}
 +\ii\delta^a_b\eta,
 \qquad
 \Omega^{11}{}_a=\frac{\ii}{2}\bar E_a,
 \qquad
 \Omega^{11}{}_{\bar a}=-\frac{\ii}{2}E_a.
\end{equation}
These formulas reproduce \rf{124}--\rf{189}, including the background
Hopf leg $E_3^{11}=1/k$ and the cancellation displayed in
\rf{191}.
\fi 


\section{Coincident propagator  constants  in  spectral $\zeta$-function   
   regularization}\label{api}

Let us first recall some  relevant  expressions for the  limits of $\AdS_2$   propagators. 
For a real massive scalar  with operator $-\nabla^2 + m^2$ one has  \cite{Camporesi:1990wm}
\begin{equation}
 \zeta_b(s;m)
 =\frac{1}{2\pi^2}\int_0^\infty\dd\nu\,
 \frac{\pi\nu\tanh(\pi\nu)}
 {\big(\nu^2+ m^2+\frac14\big)^s}.
 \label{i1}
\end{equation}
For a 2d  Majorana  fermion 
the  $\zeta$ function  for the  $-\slashed \nabla^2 + m^2$ operator  is \cite{Camporesi:1995fb}
\begin{equation}
 \zeta_f(s;m)
 =\frac{1}{2\pi^2}\int_0^\infty\dd\nu\,
 \frac{\pi\nu\coth(\pi\nu)}{(\nu^2+m^2)^s}.
 \label{i00}
\end{equation}
We may define the finite value at unit argument by
$
 \zeta(1)
 \equiv
 \big.\frac{\dd}{\dd s}
 \big[\Lambda^{2(s-1)}(s-1)\zeta(s)\big]\big|_{s=1},
 $
where $\Lambda$ is dimensionless in units of the $\AdS_2$ radius.
Then the expression in \rf{20} corresponds to 
\begin{equation}
 {
 \G_b(m)\equiv\zeta_b(1;m)
 =\tfrac{1}{2\pi}
 \big[\log\Lambda-\uppsi\big(\Delta\big)\big], \qquad \ \ \ \ \  \Delta =\tfrac12+\sqrt{m^2 +\tfrac{1}{4}}\ . }
 \label{i2}
\end{equation}
The constant in the   coincident limit of  the first-order fermion
propagator is as in \rf{020}
\be
\G_f(m) =m\,   \zeta_f(1;m) \ , \qquad \qquad 
\zeta_f(1;m)
 =\tfrac{1}{2\pi}
 \big[\log\Lambda-\uppsi(|m|)-\tfrac{1}{2|m|}\big].
 \label{i33}
\end{equation}

\vthree Let us note  that  the value of  $\G_f$  at $m=0$   can  also be obtained  directly, without  taking a limit  in \rf{i33}. 
Splitting  the propagator into the  direct  and the boundary  parts, 
$ S=S_0(x-x')+S_{\rm bdy}$   with $S_0(x)=\tfrac{1}{2\pi}\g^\mu x_\mu/x^2$, and using  $\{\g^\mu,\Gb\}=0$  one has 
${\rm tr}\,[\Gb\,S_0]=0$  identically,  so that   with the  normalization \rf{67},\rf{020} 
\be 
 \G_f(0)=\ha \,{\rm tr}\,\big[\Gb\,S_{\rm bdy}(x,x)\big] \ , 
 \la{i34a}
\ee
and then  the  $\Gb$  component  of the coincident  propagator is  {\it exactly}  boundary data  and carries no  short-distance 
divergence  (this is the  structural  reason why  the  $\log\Lambda$  in \rf{020}  comes  multiplied by $m$). 

The massless  $\AdS_2$  Dirac operator is  conformal to the flat one on the upper half plane, 
$S_{\AdS_2}(x,x')=(zz')^{1/2}S_{\mathbb H}(x,x')$, and  with the image  ansatz 
$S_{\mathbb H}=S_0(x-x')+S_0(x-\tilde x')B$,  $\tilde x'=(t',-z')$,  the boundary condition \rf{78}  requires 
$\PPi_\eta\g^{\hat t}(1+B)=0$  and $\PPi_\eta\g^{\hat z}(B-1)=0$, which are solved by  $B=-\eta \uN$. 
Then 
\be 
 S_{\rm bdy}\big|_{x'\to x} = z\cdot\frac{\g^{\hat z}}{4\pi z}\big(-\eta\epsilon_\T\,\g^{\hat z}\Gb\big)= -\frac{\eta\, \epsilon_\T}{4\pi}\,\Gb \ , 
 \qquad \qquad \qquad 
 \G_f(0)= \frac{\eta\, \epsilon_\T}{4\pi} \ , 
 \la{i34}
\ee
the magnitude  being  entirely  local, $\tfrac{1}{2\pi}\times\tfrac{1}{2z}\times z=\tfrac{1}{4\pi}$. 
Setting  $\eta={\rm sign}(\epsilon_\T m)$, 
which by \rf{78b}  is the  normalizable  domain,  reproduces  \rf{25}; this is a check on the 
conventions  rather than an input, since  no  limit  was used in deriving \rf{i34}. 

Given  the list of masses in \rf{17b}    we  may   sum over $n$ to compute also the corresponding 
propagator constants  on $\AdS_2 \times S^1$. 
The constants entering $\cX_b$ in \rf{55}   defined in  \rf{194}--\rf{1990} are ($\Phi=(x,w)$)
\begin{align}
 \G_\Phi^{(3)}=L^{-1}\sum_{n=-\infty}^\infty
 n\G_b(m_{\Phi,n}),\qquad \qquad 
 \G_\Phi^{(33)}=L^{-1}\sum_{n=-\infty}^\infty
 n^2\G_b (m_{\Phi,n}),\qquad   \ \ L^{-1} = \tfrac{k}{ 4\pi} \ , 
 \label{i3}
\end{align}
and thus 
\begin{align}\te 
&\te \G_x^{(3)}=- \tfrac1{8\pi^2}  \big(1+\frac{4\pi}{k} \cot \frac{2\pi}{k} \big), \qquad \qquad \qquad \ \ \ 
 \G_w^{(3)}=\frac1{8 \pi^2},\no
 \\
 &\te \G_x^{(33)}=- \tfrac{k}{16\pi^4}
 \big[\HH(k)+\frac{16\pi^3}{k^3}
 \cot\frac{2\pi}{k}\big],\qquad \qquad 
 \G_w^{(33)}=-\frac{k}{16\pi^4} \HH(k) \ , \la{i55} \\
& \qquad \ \ \HH(k)
 =\zeta(3)+\tfrac{4}{k^3}\int_0^\infty\dd x\,
 \frac{x^2\coth{x\ov k}}{e^x-1}=\zeta(3)\big(1+\tfrac{8}{k^3}\big)
 -\sum_{n=1}^{\infty}\uppsi''\big(1+\tfrac{1}{2}k n\big).
 \label{i5}
\end{align}
The low $k$ values are $
 \HH(1)=\frac{4\pi^2}{3},\
 \HH(2)=\frac{\pi^2}{3},\
 \HH(4)=\frac{\pi^2}{6}.
$
The  large-$k$ asymptotic expansion  of $\HH(k)$ may  be written as ($B_n$ are Bernoulli numbers, $B_0=1$) 
\begin{equation}
 \HH(k)= \zeta(3)-\tfrac14 \SS \big(\tfrac{4\pi}{k}\big),\qquad 
 \qquad\ \ \ 
  \SS (z)=\sum_{n=1}^{\infty}
 \frac{(-1)^n B_{2n}B_{2n-2}}{n(2n-2)!}\,z^{2n} \ . 
 \label{i6}
\end{equation}
The explicit form of the  combination $\cX_b$  in \rf{55}   is found to be 
(note that $\cot\frac{2\pi}{k}$ terms cancel out)
\begin{align}
 \cX_b=
\tfrac{k}{2}\Big[-2\G_x^{(3)}
 + k\G_x^{(33)}
 +3\big(2\G_w^{(3)}+k\G_w^{(33)}\big)\Big]
=\tfrac{k}{2\pi^2}
 \Big[1-\tfrac{k^2}{4\pi^2}\HH(k)\Big]
\ . 
 \label{i7}
\end{align}
A similar expression  is found for $\cX_f$ in \rf{60} with non-trivial  $\HH(k)$ terms 
 cancelling in the  combination $\cX_b-\cX_f$  in \rf{62} and leading, for $k>2$, 
  to the expression in  \rf{244}.

\section{Special cases of $k=1,2$}\label{apj}

The  cases of $k=1,2$   need  a special treatment. 
These are known to be cases of enhanced supersymmetry  leading to the appearance of extra massless
 fermionic modes in  the fluctuation  spectrum in \rf{17b} \ci{Sakaguchi:2010dg,Giombi:2023vzu}. 
 Namely,   the three triplet towers with
$m_n=p_n-1$ become massless at $n=n_*$
\begin{equation}
 n_*=2k^{-1} \ ,\qquad \ \ \ p_{n_*}=1 \ ,
 \la{313}
\end{equation}
and the three conjugate towers with
$m_n=-p_n-1$ become massless at $n=-n_*$. 
\vfour Formally approaching these zeros from the positive-mass side would
assign to all six modes the value $\G_f(0^+)=\tfrac{1}{4\pi}$
as in \rf{25},\rf{79}. This uniform assignment, however, does not
preserve the supercharge considered below.

Concretely, for the supercharge determined by \eqref{sk1}, the boundary conditions for the additional massless fermions follow from the pairings \eqref{sk4}:
\begin{equation}
 (x_{n_*},\psi^3_{n_*}) \ , \qquad
 (w^1_{-n_*},(\psi^2)^c_{-n_*}) \ ,\qquad 
 (w^2_{-n_*},-(\psi^1)^c_{-n_*})  \ .
\end{equation}
The scalar in each pair is massless and obeys the Dirichlet boundary condition. For the massless pair $(\Phi_{n}, \Psi_{n})$, with the Dirichlet scalar, the boundary expansion becomes
\begin{align}
 \Phi_n=z\Phi_{n,(1)}+\dots \ ,\qquad
 \Psi_n=z^{1/2}\Psi_{n,(0)}+\dots \ .
\end{align}
Using $\varepsilon=z^{-1/2}\varepsilon_{(0)}+\ldots$ for the Killing spinor \eqref{sk2} with
$\PPi_{+}\varepsilon_{(0)}=\varepsilon_{(0)}$, one finds from \rf{sk3}
\begin{align}
 \delta_\epsilon\Psi_{n,(0)}
 =\Phi_{n,(1)}\epsilon_{(0)} \ ,
\end{align}
as  the term $(P+q_\Psi)\Phi_n$  in $\delta_\epsilon\Psi$ in   \rf{sk3}
  vanishes at the corresponding massless level. It then follows that the boundary condition $\PPi_-\Psi_{n,(0)}=0$ is preserved by the supersymmetry and together with the boundary condition for the conjugate spinor \rf{sk5} gives
\begin{align}
 \lim_{z\to0}z^{-1/2}\PPi_-\Psi_n=0 \ ,\qquad
 \lim_{z\to0}z^{-1/2}\PPi_+\Psi^c_{-n}=0  \ .
\end{align}
Explicitly,  for the pairs above, we get   in terms of the unconjugate variables
\begin{align}
 \lim_{z\to0}z^{-1/2}\PPi_-\psi^3_{n_*}=0 \ ,\qquad
 \lim_{z\to0}z^{-1/2}\PPi_+\psi^{1,2}_{n_*}=0  \ .
\end{align}
Using \rf{773}, the corresponding coincident propagator coefficients are
\begin{equation}
\G_f(0)\Big|_{\psi^1_{n_*}}=\G_f(0)\Big|_{\psi^2_{n_*}}=\frac{1}{4\pi}\ , \qquad \G_f(0)\Big|_{\psi^3_{n_*}}=-\frac{1}{4\pi} \ . \la{314}
\end{equation}
Keeping  $\G_f(0)$  at the exceptional  level  general  for a moment,  one finds  from \rf{23},\rf{21} 
\begin{equation}
\te \X_0 = - 2\,\G_f(0)\big|_{\chi_0} =- {1\ov 2\pi} \ , \quad
 \X_{n_*}+\X_{-n_*} = - \frac{1}{2\pi} - 2\,\sum_{a=1}^{3}\G_f(0)\Big|_{\psi^a_{n_*}}  \ , \quad \big( \X_{n}+\X_{-n}\big)_{ n\ne 0,\, n_* }= - \frac{1}{\pi} . 
 \la{315}
\end{equation}
 For $k=1$ there is  also  a low-lying  pair at $p_n=\frac12$, but  one can check directly that it  gives  the same universal value $-\frac{1}{\pi}$ as in \rf{315}.\foot{The  above analysis  used the boundary conditions preserving the
supercharge specified in \rf{sk1}. Compatibility with the other
preserved supercharges, including the additional ones at $k=1,2$,
requires a separate analysis, and  should  they  impose  further 
correlations  at  $p_{n_*}=1$  the  assignment  \rf{314}  may  change.}
 
 As a result,   for $k=1$ and $k=2$ we find that, as for $k>2$,   the 
   combination  $\cX$ in \rf{22} and thus the 2-loop coefficient 
$f_2$   vanish. 
Let us  recall  that 
the results   for the 1-loop M2 brane partition function  \cite{Giombi:2023vzu}  are
 different for different $k$:
$Z_1(1) = {1\ov 4}; \   Z_1(2) = 1; \   Z_1( k>2) = (2\sin{2\pi\ov k})^{-1}$.  

We should  add a warning that while  for $k>2 $   the classical M2 brane solution localized at a point  in $\CP^3$ and 
wrapped on Hopf cycle is stable,  for $k=1,2$ there should  be a continuous   family of degenerate BPS saddles
that  should be accounted for.  We ignored  this  circumstance  above  as it requires a  separate  treatment beyond the goals of the present paper. 
Let us note  also  that a  localization  prediction  for the ABJM  Wilson loop expectation value  
 is currently unavailable  for $k=1,2$.


\iffa 
\foot{The origin of the nonvanishing result at $k=1,2$ can be isolated precisely.
At the exceptional pair $p_{n_*}=1$ both fermionic and bosonic masses cross
zero, but only the boson one   leads to a change.
 On the fermionic side the six
massless triplets enter through the combination $(a-1)Z_F(a-1)$, a
$0\times\infty$ limit that the boundary-condition analysis of
Section~\ref{sec:X-result} resolves to the one-sided value
$S(0^+)=\frac1{4\pi}$ \rf{314}; with this assignment the exceptional
fermionic pair \rf{316} coincides exactly with the analytic continuation of
the generic formula \rf{307}, so the fermions produce no deviation at all.
On the bosonic side the charged $x$ mode at $n=n_*$ is massless with tower
dimension $\Delta_{x}=a-1\to0$ \rf{305}: the closed pair formula \rf{306}
was derived by shifting $\psi(a-1)=\psi(a)-\frac1{a-1}$ along this tower,
and at $a=1$ it crosses the infrared branch point of $\psi(\Delta)$ in the
coincident propagator \rf{298}, so that the vanishing charge weight
$2a(a-1)=2a\Delta_x$ times the pole leaves the spurious finite residue
$\lim_{a\to1}\frac{a(a-1)}{\pi(a-1)}=\frac1\pi$ in the continued formula.
The direct lattice evaluation \rf{315} contains no such term --- at $a=1$
the massless $x$ decouples identically, since its weight is exact charge
kinematics of the fibre-current composite, not a limit to be prescribed ---
and is therefore smaller by $\frac1\pi$ (the $-9$ in place of $-8$); the
accompanying massless $w$ at $n=-n_*$ is harmless, as it sits on the
regular $\Delta=1$ branch with finite $\psi(1)$.  This purely bosonic
mismatch is the entire exceptional deviation
$X_{n_*}+X_{-n_*}=-\frac2\pi$ \rf{317} from the universal $-\frac1\pi$, and
on the physical branch, where the $n=0$ level already matches the universal
value, it is all that survives the sum:
$\mathscr X_{k=1,2}=-\frac{\nu_k}{\pi}=-\frac{k}{4\pi^2}$, hence
$f_2=\frac{3k}{2\pi^2}$.}
\fi

{\small
\bibliographystyle{JHEP-v2.9}
\bibliography{m2s}
}

\end{document}